\documentclass[11pt]{article}
\usepackage{graphicx}
\usepackage{amssymb}
\usepackage{amsmath}
\usepackage{graphicx}
\usepackage{amstext}
\usepackage{esint}
\usepackage[utf8]{inputenc}
\usepackage{color}
\usepackage{cite}

\def\beq{\begin{eqnarray}}    %%%  begequation/eqnarray
\def\eeq{\end{eqnarray}}      %%%  endequation/eqnarray
\newcommand{\rL}{\rho_\Lambda}

\newcommand{\CC}{\Lambda}

\begin{document}
%\pubblock

%\today

%\vspace{1cm}

 \hyphenation{nu-cleo-syn-the-sis u-sing si-mu-la-te ma-king
cos-mo-lo-gy know-led-ge e-vi-den-ce stu-dies be-ha-vi-or
res-pec-ti-ve-ly appro-xi-ma-te-ly gra-vi-ty sca-ling
ge-ne-ra-li-zed re-mai-ning pa-ra-me-ter}

%%%%%%%%%%%%%%%%%%%%%%%%%%%%%%%%%%%%%%%%%%%%%%%%%%%%%%%%%

%\newpage

%%%%%%%%%%%%%%%%%%%%%%%%%%%%%%%%%%%%%%%%%%%%%%%%%%%%%%%%%
%\flushright{UB-ECM-PF-06/24 }\\

\begin{center}
{\it\LARGE Signs of Dynamical Dark Energy in Current Observations} \vskip 2mm

 \vskip 8mm

\textbf{\large Joan Sol\`a Peracaula, Adri\`a G\'omez-Valent and Javier de Cruz P\'erez}

\vskip 0.5cm
Departament de F\'isica Qu\`antica i Astrof\'isica, and Institute of Cosmos Sciences,\\ Universitat de Barcelona, \\
Av. Diagonal 647, E-08028 Barcelona, Catalonia, Spain

\vskip0.5cm

\vskip0.4cm

E-mails: sola@fqa.ub.edu, adriagova@fqa.ub.edu, decruz@fqa.ub.edu

 \vskip2mm

\end{center}
\vskip 15mm

\begin{quotation}
\noindent {\large\it \underline{Abstract}}.
Investigations on dark energy (DE) are currently inconclusive about its time evolution. Hints of this possibility do however glow now and then in the horizon.  Herein we assess the current status of dynamical dark energy (DDE) in the light of a large body of updated  SNIa+$H(z)$+BAO+LSS+CMB observations, using the full Planck 2015 CMB likelihood.
The performance of the $\CC$CDM model (with equation of state $w=-1$ for $\CC$) is confronted with that of the general XCDM and CPL parametrizations, as well as with the traditional $\phi$CDM model based on the scalar field potential $V\sim \phi^{-\alpha}$. In particular, we gauge the impact of  the bispectrum in the LSS and BAO parts, and show that the subset of CMB+BAO+LSS observations may contain the bulk  of the DDE signal. The departure from $w=-1$ is significant: roughly $2.6\sigma$  for XCDM and $2.9\sigma$ for $\phi$CDM. In both cases the full Bayesian evidence is found to be positive even for  a prior range of the DDE parameters extending over several standard deviations from the mean when the bispectrum is taken into account.  Positive signs follow as well from the preliminary results of Planck 2018 data using the compressed CMB likelihood.  As a bonus we also find that the  $\sigma_8$-tension becomes reduced with DDE.
\end{quotation}
\vskip 5mm

\begin{quotation}

\end{quotation}

\newpage

%\tableofcontents

\newpage

%%%%%%%%%%%%%%%%%%%%%%%%%%%%%%%%%%%%%%%%%%%%%%%%%%%%%%%%%%%%%%%%%
%%%%%%%%%%%%%%%%%%%%%%%%%%%%%%%%%%%%%%%%%%%%%%%%%%%%%%%%%%%%%%%%%
%%%%%%%%%%%%%%%%%%%%%%%%%%%%%%%%%%%%%%%%%%%%%%%%%%%%%%%%%%%%%%%%%

\section{Introduction}\label{intro}

Even if the existence proof of the DE is not iron-clad, the great majority of cosmologists are agreed that the universe is in accelerated expansion\,\cite{SNIaRiess,SNIaPerlmutter,Planck}  and that some physical cause must be responsible for it.  The canonical picture in GR is to assume that such cause is the presence of a cosmological constant (CC) term, $\CC$, in Einstein's equations. No reason, however, is given for its constancy throughout the entire cosmic history. In fact, $\CC=$const. is not required by the cosmological principle. Such oversimplification might be at the root of the Cosmological Constant Problem, namely the appalling discrepancy between the measured value of the vacuum energy density,  $\rL=\CC/(8\pi G)\sim 10^{-47}$ GeV$^4$ (G being Newton's constant), and the incommensurably larger value predicted in quantum field theory (QFT)\,\cite{Weinberg,PeeblesRatra2003,Padmanabhan2003,JSPRev2013}. The cosmological constant problem stays unsolved right now, even after more than a full century after $\CC$ was first introduced by Einstein. Most likely the mystery will remain for a long time still. In this work, we will not attempt to solve it. Our main aim is much more humble. Somehow we wish to follow the original phenomenological approach that made possible to unveil that $\CC$ is nonvanishing, irrespective of its ultimate physical nature. The method to substantiate that $\CC$ is non-null was largely empirical, namely $\CC$ was assumed to be a parameter and then fitted directly to the data. Of course a minimal set of assumptions had to be made, such as the validity of the Cosmological Principle and hence of the Friedmann-Lema\^\i tre-Robertson-Walker (FLRW) metric, with the ensuing set of Friedmann equations for the scale factor.  In our case, we wish of course to keep these minimal assumptions and make a phenomenological case study of the next-to-leading possibility, which is that the $\CC$, or in general the DE, might be not just a parameter but a slowly varying cosmic variable that stays close to a constant for cosmic spans of time and therefore capable of  mimicking the $\CC$CDM-like behavior.

This is all the most interesting if we {take into account that, aside from theoretical problems, a number of persisting tensions with the data (particularly on the local value of the Hubble parameter\,\cite{Riess2016,Riess2018} and the large scale structure  formation data\,\cite{Macaulay2013})  suggest  that the concordance or $\CC$CDM model, with rigid $\CC$-term, might be performing insufficiently at the observational level. The possibility to alleviate some of these problems by assuming that $\rL$, or in general the DE, might be (slowly) dynamical is a  natural option to test, see  e.g. \cite{Valentino2017,PLB2017,AdriaJoan2017}. Here we are primarily spurred by these phenomenological considerations and the disparity of different results. For instance, hints of DDE were highlighted in\,\cite{Salvatelli2014,Sahni2014,ApJL2015}. In addition, nonparametric studies of the DDE suggest positive evidence at about $3.5\sigma$ c.l., see \,\cite{GBZhao2017}. Such confidence level is actually comparable to that achieved through specific models of DDE,  in which attempts were made to cure the mentioned tensions -- see e.g. \,\cite{ApJ2017,IJMP,EPL,MNRAS2018} and references therein. Other recent works, in contrast, find a variety of levels of evidence in different contexts  or simply no sizeable deviation from the $\CC$CDM, see \cite{Ferreira2017,Costa2017,Li2016,Heavens2017PRL,Ooba2017,ParkRatra_data,ParkRatraXCDM,ParkRatraPhiCDM,Ooba2018,ParkRatra2018,Tsiapi2018,Dutta2018,
Martinelli2019} and references therein, for example.

In this paper, we reassess the current situation on the DDE and try to clarify why some authors find evidence whereas others find no evidence at all. The novelty is that we pay due attention to the sensitivity of the signal to potentially relevant features sitting in the baryon acoustic oscillations (BAO) and large-scale structure (LSS) formation data which have not been explored in previous studies. Most noticeably we explore  the inclusion of the matter bispectrum together with the power spectrum. We find that this ingredient helps to develop more sensitivity to the DDE signal.  We test and confirm this fact by considering the reaction of well-known generic parametrizations of the DDE, such as the XCDM and CPL, as well as a well-known representative of the traditional class of scalar field models ($\phi$CDM). Our analysis is performed on the basis of a large set of updated  SNIa+$H(z)$+BAO+LSS+CMB observations and leads to a signal of DDE in the ballpark of $3\sigma$ c.l. Such positive test is encouraging. If revalidated with the advent of future data, it could guide us into  future strategies  helping to consolidate  the DDE signature.

%
%%%%%%%%%%%%%%%%%%%%%%%%%%%%%%%%%%%%%%%%%%%%%%%%%%%%%%%%%%%%%%%%%%
%%%%%%%%%%%%%%%%%%%%%%%%%%%%%%%%%%%%%%%%%%%%%%%%%%%%%%%%%%%%%%%%%
%%%%%%%%%%%%%%%%%%%%%%%%%%%%%%%%%%%%%%%%%%%%%%%%%%%%%%%%%%%%%%%%%

\renewcommand{\arraystretch}{1.1}
\begin{table}[t!]
\begin{center}
\resizebox{1\textwidth}{!}{

\begin{tabular}{|c  ||c | c | c || c | c |c | c |c|}
 \multicolumn{1}{c}{} & \multicolumn{3}{c}{DS1 with Spectrum (DS1/SP)} & \multicolumn{4}{c}{DS1 with Bispectrum (DS1/BSP)}
\\\hline
{\scriptsize Parameter} & {\scriptsize $\Lambda$CDM} & {\scriptsize XCDM} & {\scriptsize $\phi$CDM} & {\scriptsize $\Lambda$CDM} & {\scriptsize XCDM} & {\scriptsize CPL} & {\scriptsize $\phi$CDM}
\\\hline
{\scriptsize $H_0$ (km/s/Mpc)} & {\scriptsize $69.22^{+0.48}_{-0.49}$} & {\scriptsize $68.97^{+0.76}_{-0.79}$} & {\scriptsize $68.70^{+0.66}_{-0.61}$} & {\scriptsize $68.21^{+0.40}_{-0.38}$} & {\scriptsize $67.18^{+0.63}_{-0.68} $} & {{\scriptsize $67.17\pm 0.72$}} & {{\scriptsize $67.19^{+0.67}_{-0.64} $}}
\\\hline
$\omega_{cdm}$ & {\scriptsize $0.1155^{+0.0011}_{-0.0010}$} & {\scriptsize $0.1151^{+0.0013}_{-0.0014}$} & {\scriptsize $0.1147^{+0.0013}_{-0.0012}$} & {\scriptsize $0.1176^{+0.0008}_{-0.0009}$} & {\scriptsize $0.1161\pm 0.0012$} & {{\scriptsize $0.1161^{+0.0015}_{-0.0014}$}} & {\scriptsize $0.1160^{+0.0013}_{-0.0012}$}
\\\hline
$\omega_{b}$ & {{\scriptsize$0.02247^{+0.00020}_{-0.00019}$}} & {{\scriptsize$0.02251^{+0.00021}_{-0.00020}$}} & {{\scriptsize$0.02255^{+0.00020}_{-0.00022}$}} & {\scriptsize$ 0.02231^{+0.00019}_{-0.00018}$} & {{\scriptsize$0.02244^{+0.00021}_{-0.00020}$}} & {{\scriptsize$0.02244\pm 0.00021$}} &  {{\scriptsize $0.02245\pm 0.00021$}}
\\\hline
$\tau$ & {{\scriptsize$0.079^{+0.012}_{-0.013}$}} & {{\scriptsize$0.084^{+0.014}_{-0.017}$}} & {{\scriptsize$0.088\pm 0.016$}} & {{\scriptsize$0.059^{+0.012}_{-0.009}$}} & {{\scriptsize$0.074^{+0.013}_{-0.015}$}} & {{\scriptsize$0.074^{+0.016}_{-0.017}$}} &  {{\scriptsize$0.074^{+0.014}_{-0.016}$}}
\\\hline
$n_s$ & {{\scriptsize$0.9742\pm 0.0042$}} & {{\scriptsize$0.9752^{+0.0048}_{-0.0052}$}} & {{\scriptsize$0.9768\pm 0.0048$}} & {{\scriptsize$0.9678\pm 0.0038$}} & {{\scriptsize$0.9724\pm 0.0047$}} &  {{\scriptsize$0.9724^{+0.0049}_{-0.0052}$}} &  {{\scriptsize$0.9729^{+0.0044}_{-0.0048}$}}
\\\hline
$\sigma_8(0)$  & {{\scriptsize$0.813^{+0.008}_{-0.009}$}} & {{\scriptsize$0.811\pm 0.010$}} & {{\scriptsize$0.808\pm 0.010$}} & {\scriptsize $0.804^{+0.007}_{-0.008}$} & {{\scriptsize$0.795^{+0.010}_{-0.009}$}} & {{\scriptsize$0.795 \pm 0.010$}} &  {{\scriptsize$0.793\pm 0.009$}}
\\\hline
$w_0$ & {\scriptsize -1} & {{\scriptsize $-0.986^{+0.030}_{-0.029}$}} & - & {\scriptsize -1} &  {{\scriptsize$-0.945^{+0.029}_{-0.028}$}} &  {{\scriptsize$-0.934^{+0.067}_{-0.075}$}} & -
\\\hline
$w_1$ & - & - & - & - &  - & {{\scriptsize $-0.045^{+0.273}_{-0.204}$}} & -
\\\hline
$\left(\alpha,  10^{-3}\bar{\kappa}\right)$ & - & - & { \scriptsize $\left(<0.092, 37.3^{+1.9}_{-2.3}\right)$ } & - &  - & - & {\scriptsize $\left(0.150^{+0.070}_{-0.086}\right.$}, {\scriptsize $\left.33.5^{+1.1}_{-2.1}\right)$}
\\\hline
\end{tabular}}
\end{center}
\label{tableFit1}
\caption{\scriptsize The mean fit values and $68.3\%$ confidence limits for  the considered models using dataset DS1, i.e.  all SNIa+$H(z)$+BAO+LSS+CMB data, with full Planck 2015 CMB likelihood.  In all cases a massive neutrino of  $0.06$ eV has been included. The first block involves BAO+LSS data using the matter (power) spectrum (SP) and is  labelled  DS1/SP. The second block includes both  spectrum and bispectrum, and is denoted DS1/BSP (see text). We display the fitting results for the relevant parameters, among them those that characterize the DDE models under discussion: the EoS parameter $w_0$ for XCDM, $w_0$ and $w_1$ for the CPL, the power $\alpha$ of the potential and $\bar{\kappa}\equiv\kappa [{\rm M_P/(km/s/Mpc)}]^2$ for $\phi$CDM, as well as the six conventional parameters:  the Hubble parameter $H_0$, $\omega_{cdm}=\Omega_{cdm} h^2$ and $\omega_{b}=\Omega_{b} h^2$ for cold dark matter and baryons, the reionization optical depth $\tau$, the spectral index $n_s$ of the primordial power-law power spectrum, and, for convenience, instead of the amplitudes $A_s$ of such spectrum we list the values of $\sigma_8(0)$.}
\end{table}

\section{DDE parametrizations and models}\label{sect:DDEmodels}

We  consider two generic parametrizations of the DDE, together with a well-known  $\phi$CDM model, and confront them to a large and updated set of SNIa+$H(z)$+BAO+LSS+CMB observations.  Natural units are used hereafter, although we keep explicitly Newton's $G$, or equivalently the Planck mass:  $M_P = 1/\sqrt{G} = 1.2211\times 10^{19}$ GeV. Flat FLRW metric is  assumed throughout: $ds^2=-dt^2+a^2(t)(dx^2+dy^2+dz^2)$, where $a(t)$ is the scale factor as a function of the cosmic time.

\subsection{XCDM and CPL}

The first of the DDE parametrizations under study is the conventional XCDM\,\cite{XCDM}. In it both matter and DE are self-conserved (non-interacting) and the DE density is simply given by $\rho_{X}(a) = \rho_{X0}a^{-3(1+w_0)}$, where $\rho_{X0} = \rho_\Lambda$ is the current value and $w_0$ the (constant)  equation of state (EoS) parameter of the DE fluid.
For $w_0=-1$ we recover the $\Lambda$CDM model with a rigid CC. For $w_0 \gtrsim-1$ the XCDM mimics quintessence, whereas for $w_0 \lesssim -1$ it mimics phantom DE.
It is worth checking if a dynamical EoS for the DE can furnish a better description of the observational data. So we consider the well-known CPL parametrization\,\cite{CPL,Linder}, which is characterized by the following  EoS:

%
%%%%%%%%%%%%%%%%%%%%%%%%%%%%%%%%%%%%%%%%%%%%%%%%%%%%%%%%%%%%%%%%%%%%%%%%%%%%%%%%%%
\renewcommand{\arraystretch}{1.1}
\begin{table}[t!]
\begin{center}
\resizebox{1\textwidth}{!}{

\begin{tabular}{|c  ||c | c | c || c | c |c |}
 \multicolumn{1}{c}{} & \multicolumn{3}{c}{DS2 with Bispectrum  (DS2/BSP)} & \multicolumn{3}{c}{DS2/BSP with Planck 2018 (compressed likelihood)}
\\\hline
{\scriptsize Parameter} & {\scriptsize $\Lambda$CDM} & {\scriptsize XCDM}& {\scriptsize $\phi$CDM} & {\scriptsize $\Lambda$CDM} & {\scriptsize XCDM} & {\scriptsize $\phi$CDM}
\\\hline
{\scriptsize $H_0$ (km/s/Mpc)} & {\scriptsize $68.20^{+0.38}_{-0.41}$} & {\scriptsize $66.36^{+0.76}_{-0.86} $} & {\scriptsize $66.45\pm 0.74 $} & {\scriptsize $68.92\pm 0.31 $} & {\scriptsize $67.08\pm 0.73 $} & {\scriptsize $67.02\pm 0.70 $}
\\\hline
$\omega_{cdm}$ & {{\scriptsize$0.1176^{+0.0008}_{-0.0009}$}} & {{\scriptsize$0.1155^{+0.0014}_{-0.0012}$}} & {{\scriptsize$0.1154\pm 0.0013$}} & {{\scriptsize$0.1197\pm 0.0007$}} & {{\scriptsize$0.1191\pm 0.0008$}} & {{\scriptsize$0.1191\pm 0.0008$}}
\\\hline
$\omega_{b}$ & {{\scriptsize$0.02230\pm 0.00019$}} & {{\scriptsize$0.02247\pm 0.00021$}} & {{\scriptsize$0.02248^{+0.00020}_{-0.00021}$}} & {{\scriptsize$0.02254\pm 0.00013$}} & {{\scriptsize$0.02259\pm 0.00013$}} & {{\scriptsize$0.02259\pm 0.00013$}}
\\\hline
$\tau$ & {{\scriptsize$0.059\pm 0.010$}} & {{\scriptsize$0.083^{+0.014}_{-0.018}$}} & {{\scriptsize$0.083^{+0.013}_{-0.014}$}} & - & - & -
\\\hline
$n_s$ & {{\scriptsize$0.9681\pm 0.0039$}} & {{\scriptsize$0.9742^{+0.0048}_{-0.0053}$}} & {{\scriptsize$0.9746^{+0.0044}_{-0.0051}$}} & {{\scriptsize$0.9727\pm 0.0033$}} & {{\scriptsize$0.9734\pm 0.0032$}} & {{\scriptsize$0.9735\pm 0.0032$}}
\\\hline
$\sigma_8(0)$  & {{\scriptsize$0.805\pm 0.007$}} & {{\scriptsize$0.788\pm 0.010$}} & {{\scriptsize$0.789\pm 0.010$}} & {{\scriptsize$0.800\pm 0.008$}} & {{\scriptsize$0.774\pm 0.013$}} & {{\scriptsize$0.773\pm 0.012$}}
\\\hline
$w_0$ & {\scriptsize -1} & {{\scriptsize$-0.911^{+0.035}_{-0.034}$}} & - & {\scriptsize -1} & {{\scriptsize$-0.932\pm 0.025$}} &  -
\\\hline
$\left(\alpha,  10^{-3}\bar{\kappa}\right)$ & - & - & {{\scriptsize$\left(0.240^{+0.086}_{-0.102}\right.$}, {\scriptsize $\left.32.0^{+0.8}_{-1.4}\right)$}} & - & - &  {\scriptsize $\left(0.168\pm 0.068\right.$}, {\scriptsize $\left.32.7\pm 1.5\right)$}
\\\hline
\end{tabular}}
\end{center}
\label{tableFit1}
\caption{\scriptsize As in Table 1, but using dataset DS2/BSP, which involves BAO+LSS+CMB data only. In the first block we use the full likelihood for Planck 2015, whereas in the second we use the compressed CMB likelihood for the more recent Planck 2018  data. See text for more details.}
\end{table}
%%%%%%%%%%%%%%%%%%%%%%%%%%%%%%%%%%%%%%%%%%%%%%%%%%%%%%%%%%%%%%%%%%%%%%%%%%%%%%%%%%

%
\begin{equation}\label{eq:CPL}
w(a) = w_0 + w_1(1-a) = w_0 + w_1\frac{z}{1+z}\,,
\end{equation}
where $z=a^{-1}-1$ is the cosmological redshift.
The corresponding Hubble rate $H=\dot{a}/a$ normalized with respect to the current value, $H_0=H(a=1)$, takes the form:
\begin{equation}
E^2(a)\equiv\frac{H^2(a)}{H^2_0}=(\Omega_b+\Omega_{cdm}){a^{-3}} + \Omega_\gamma{a^{-4}}+ \frac{\rho_\nu(a)}{\rho_{c 0}} + \Omega_\Lambda{a^{-3(1+w_0+w_1)}}e^{-3w_1(1-a)}.
\end{equation}
Here $\Omega_i={\rho_{i 0}}/{\rho_{c 0}}$ are the current energy densities of baryons, cold dark matter, photons and CC/DE normalized with respect to the present critical density $\rho_{c 0}$.
The neutrino contribution,  $\rho_\nu(a)$ is more complicated since it contains a massive component, $\rho_{\nu, m}(a)$, apart from the massless ones. During the expansion of the universe, the massive neutrino transits from a relativistic to a nonrelativistic regime.  This process is nontrivial and has to be solved numerically.  In the above expression, for $w_1=0$ we recover the XCDM, and also setting   $w_0=-1$ we are back to the $\CC$CDM.

%%%%%%%%%%%%%%%%%%%%%%%%%%%%%%%%%%%%%%%%%%%%%%%%%%%%%%%%%%%%%%%%%
%%%%%%%%%%%%%% FIGURE 1 %%%%%%%%%%%%%%%%%%%%%%%%%%%
%%%%%%%%%%%%%%%%%%%%%%%
%%%%
\begin{figure}
\begin{center}
\label{FigContour}
\includegraphics[width=4.3in, height=2.4in]{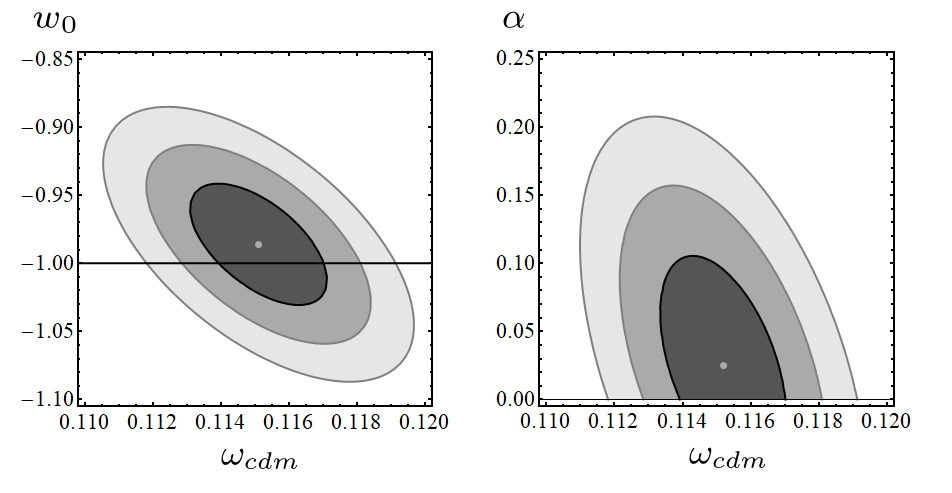}
\caption{\scriptsize {Likelihood contours for the XCDM  parametrization (left) and the considered $\phi$CDM model  (right) in the relevant planes $(\omega_{cdm},w_0)$ and $(\omega_{cdm},\alpha)$, respectively, after marginalizing over the remaining parameters. Dataset DS1/SP  is used in both cases (cf.  first block of Table 1). The various contours correspond to $1\sigma$, $2\sigma$ and $3\sigma$ c.l.}}
\end{center}
\end{figure}
%%%%%%
%%%%%%%%%%%%%%%%%%%%%%%%%%%%%%%%%%%%%%%%%%%%%%%%%%%%%%%%%%%%%%%%%
%%%%%%%%%%%%%%%%%%%%%%%%%%%%%%%%%%%%%%%%%%%%%%%%%%%%%%%%%%%%%%%%%

\subsection{$\phi$CDM}

Let us now briefly summarize the theoretical framework of the $\phi$CDM, which has a well-defined local Lagrangian description. The DE is described here in terms of a scalar field, $\phi$, which we take dimensionless. Such field is minimally coupled to curvature ($R$) and the generic action is just the sum of the Einstein-Hilbert action $S_{\rm EH}$, the scalar field action $S_\phi$, and the matter action $S_m$:
\begin{equation}\label{eq:PhiCDMLagrangian}
S=S_{\rm EH}+S_\phi+S_m=\frac{M_P^2}{16\pi}\int d^4 x \,\sqrt{-g}\,\left[R-\frac12\,g^{\mu\nu}\,\partial_{\mu}\,\phi\,\partial_\nu\phi- V(\phi)\right]+S_m\,.
\end{equation}
The energy density and pressure of $\phi$ follow from its energy-momentum tensor, $T^{\phi}_{\mu\nu}=\frac{-2}{\sqrt{-g}}\frac{\delta S_{\phi}}{\delta g^{\mu\nu}}$, and the fact that $\phi$ is a homogeneous scalar field which depends only on the cosmic time. Thus,
\begin{equation}\label{eq:densitypressure}
\rho_\phi=T^{\phi}_{00} = \frac{M^2_P}{16\pi}\left( \frac{\dot{\phi}^2}{2} + V(\phi)\right) \quad p_\phi =T^{\phi}_{ii}= \frac{M^2_P}{16\pi}\left( \frac{\dot{\phi}^2}{2} - V(\phi)\right).
\end{equation}
Dots indicate derivatives with respect to the cosmic time, and notice that within our conventions $V(\phi)$ has dimension 2 in natural units.   %The energy scale is set by the Planck mass.

The field equation for $\phi$ is just the Klein-Gordon equation in curved spacetime, $\Box\phi+\partial V/\partial\phi=0$, which for the FLRW metric leads to
\begin{equation}\label{eq:KleinGordon}
\ddot\phi+3H\dot{\phi}+\frac{\partial V}{\partial\phi}=0\,.
\end{equation}
The corresponding Einstein's field equations read:
\begin{eqnarray}
&&3H^2=8\pi\,G\,(\rho_m+\rho_r+\rho_\phi)\label{eq:FriedmannEq}\\
&&3H^2+2\dot{H}=-8\pi\,G\,(p_m+p_r+p_\phi)\label{eq:PressureEq}\,.
\end{eqnarray}
Here $\rho_m=\rho_b+\rho_{cdm}+\rho_{\nu,m}$  involves the pressureless contributions from baryons and cold dark matter as well as the massive neutrino contribution. The latter evolves during the cosmic expansion from the relativistic regime (where $p_m=p_{\nu,m}\neq0$) to the nonrelativistic one (where $p_{\nu,m}\simeq 0$).
On the other hand, $p_r=\rho_r/3$ is the purely relativistic part from photons and the massless neutrinos.

%%%%%%%%%%%%%%%%%%%%%%%%%%%%%%%%%%%%%%%%%%%%%%%%%%%%%%%%%%%%%%%%%
%%%%%%%%%%%%%% FIGURE 1 %%%%%%%%%%%%%%%%%%%%%%%%%%%
%%%%%%%%%%%%%%%%%%%%%%%
%%%%
\begin{figure}
\begin{center}
\label{FigContour}
\includegraphics[width=4.3in, height=2.4in]{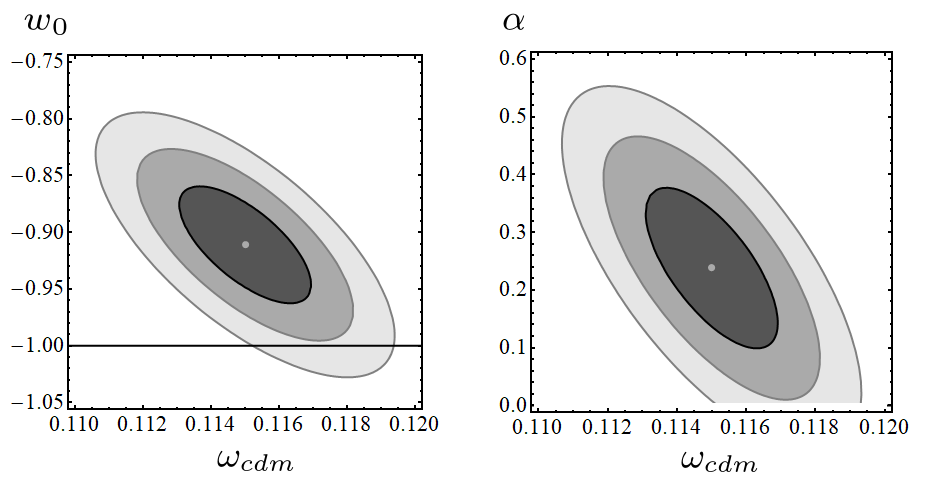}
\caption{\scriptsize{As in Fig. 1, but using dataset DS2/BSP. The various contours correspond, again, to $1\sigma$, $2\sigma$ and $3\sigma$ c.l.
The central values in both cases are shifted  $>2.5\sigma$ away from the $\CC$CDM, i.e. from $w_0 = -1$ and $\alpha=0$ in each case. Marginalization also over $\omega_{cdm}$ increases the c.l. up to $2.6\sigma$ and $2.9\sigma$, respectively  (cf.  first block of Table 2). }}
\end{center}
\end{figure}
%%%%%%
%%%%%%%%%%%%%%%%%%%%%%%%%%%%%%%%%%%%%%%%%%%%%%%%%%%%%%%%%%%%%%%%%
%%%%%%%%%%%%%%%%%%%%%%%%%%%%%%%%%%%%%%%%%%%%%%%%%%%%%%%%%%%%%%%%%

As a representative potential for our analysis we borrow the traditional quintessence potential by Peebles and Ratra\,\cite{PeeblesRatra1988,RatraPeebles1988},

\begin{equation}\label{eq:PRpotential}
%V(\phi) = \kappa\bar{H}^2_0\phi^{-\alpha},
V(\phi) = \frac12\kappa M_P^2\phi^{-\alpha}\,,
\end{equation}
where $\kappa$ is a dimensionless parameter. In this context, the value of the CC at present ($t=t_0$) is given by $\rL=\rho_\phi(t_0)$. We expect $\kappa$  to be positive since we know that $\rho_\Lambda>0$ and the potential energy at present is dominant. In what follows, when referring to the $\phi$CDM model we will implicitly assume this form of the potential.  The power $\alpha$ in it should be positive as well, but sufficiently small so that $V(\phi)$ mimics an approximate CC slowly evolving with time.  In our case we take  $\alpha$  as a free parameter while $\kappa$ is a derived one, as discussed below.

The Klein-Gordon equation with the potential (\ref{eq:PRpotential}) can be  written in terms of the scale factor as follows (prime stands for $d/da$, and we use $d/dt=a H d/da$):
\begin{equation}\label{eq:KGa}
\phi^{\prime\prime}+\left(\frac{{H}^\prime}{{H}}+\frac{4}{a}\right) \phi^\prime-\frac{\alpha}{2}\frac{{\kappa{M^2_P}}\phi^{-(\alpha+1)}}{(a {H})^2}=0\,.
\end{equation}
When the cosmic evolution is characterized by the dominant energy density  $\rho(a)\propto {a}^{-n}$ ($n=3$ for the matter-dominated epoch and $n=4$ for radiation-dominated), i.e. when the DE density is negligible, one can search for power-law solutions of the Klein-Gordon equation, i.e. solutions of the form $\phi\propto a^p$.
However,  around the current time the effect of the potential is significant and the solution has to be computed numerically.
Equation (\ref{eq:KGa}) is coupled to the cosmological equations under appropriate initial conditions. We can set these conditions at the radiation epoch, where we can neglect the DE. We find
\begin{equation}\label{eq:Initialphi}
\phi(a) = \left[\frac{3\alpha(\alpha +2)^2 \kappa M_P^4}{64\pi (\alpha +6)}\ \left(\rho_r(a)+\rho_{\nu,m}\right(a))^{-1}\right]^{1/(\alpha +2)}\,.
\end{equation}
Notice that $\rho_{r,\rm tot}\equiv\rho_r+\rho_{\nu,m}$ is the total radiation energy density, which includes also the massive neutrino contribution. At the radiation epoch, however, $\rho_{r,\rm tot}$ behaves very approximately as $\sim a^{-4}$. Thus, Eq.(\ref{eq:Initialphi}) together with $\phi'(a)$ obtained from it can be used for the numerical solution of Eq.\,(\ref{eq:KGa}) and of the EoS  of the scalar field at any subsequent epoch: $w_{\phi}(a)=p_{\phi}(a)/\rho_{\phi}(a)$.

To compare the theoretical predictions of the different models under study with the available observational data we have made use of the Einstein-Boltzmann code CLASS \cite{Blas2011} in combination with the powerful Monte Carlo Markov chain (MCMC) sampler MontePython  \cite{Audren2013}. In the particular instance of $\phi$CDM, we have conveniently modified CLASS such as to implement the shooting method, see \,\cite{StoerBulirch} for a detailed exposition, what allowed us to consistently determine the value of $\kappa$ for each value of the free parameter $H_0$. As it is well-known, such numerical technique consists in replacing a boundary value problem with an initial value one through a large number of iterations of the initial conditions until finding the optimized solution. For the $\phi$CDM, the initial conditions are determined by (\ref{eq:Initialphi}) and its time derivative, which are both explicitly dependent on $\kappa$.

%%%%%%%%%%%%%%%%%%%%%%%%%%%%%%%%%%%%%%%%%%%%%%%%%%%%%%%%%%%%%%%%%%%
%
\renewcommand{\arraystretch}{1.1}
\begin{table}[t!]
\begin{center}
%\resizebox{1\textwidth}{!}{

\begin{tabular}{|c  ||c | c | c |}
 \multicolumn{1}{c}{} & \multicolumn{3}{c}{DS3 with Spectrum  (DS3/SP)}
\\\hline
{\scriptsize Parameter} & {\scriptsize $\Lambda$CDM} & {\scriptsize XCDM} & {\scriptsize $\phi$CDM}
\\\hline
{\scriptsize $H_0$ (km/s/Mpc)} & {\scriptsize $69.30^{+0.53}_{-0.52}$} & {\scriptsize $69.26^{+0.80}_{-0.79} $} & {\scriptsize $68.80^{+0.65}_{-0.58} $}
\\\hline
$\omega_{cdm}$ & {{\scriptsize$0.1154\pm 0.0011$}} & {{\scriptsize$0.1156^{+0.0014}_{-0.0015}$}} & {{\scriptsize$0.1147^{+0.0014}_{-0.0011}$}}
\\\hline
$\omega_{b}$ & {{\scriptsize$0.02249^{+0.00019}_{-0.00021}$}} & {{\scriptsize$0.02247\pm 0.00021$}} & {{\scriptsize$0.02254^{+0.00019}_{-0.00022}$}}
\\\hline
$\tau$ & {{\scriptsize$0.086^{+0.012}_{-0.014}$}} & {{\scriptsize$0.085^{+0.015}_{-0.016}$}} & {{\scriptsize$0.093^{+0.013}_{-0.014}$}}
\\\hline
$n_s$ & {{\scriptsize$0.9750^{+0.0045}_{-0.0044}$}} & {{\scriptsize$0.9746^{+0.0051}_{-0.0050}$}} & {{\scriptsize$0.9769^{+0.0044}_{-0.0049}$}}
\\\hline
$\sigma_8(0)$  & {{\scriptsize$0.819^{+0.009}_{-0.010}$}} & {{\scriptsize$0.818\pm 0.011$}} & {{\scriptsize$0.813\pm 0.010$}}
\\\hline
$w_0$ & {\scriptsize -1} & {{\scriptsize$-1.002^{+0.034}_{-0.035}$}} & -
\\\hline
$\left(\alpha,  10^{-3}\bar{\kappa}\right)$ & - & - & { \scriptsize $\left(<0.081, 37.6\pm 2.1\right)$ }
\\\hline
\end{tabular}
%}
\end{center}
\label{tableFit1}
\caption{\scriptsize As in Tables 1 and 2, but using dataset DS3/SP. The latter is similar to the one employed in the analyses of these models in \cite{ParkRatraXCDM,ParkRatraPhiCDM}, see the text for further details.
}
\end{table}
%%%%%%%%%%%%%%%%%%%%%%%%%%%%%%%%%%%%%%%%%%%%%%%%%%%%%%%%%%%%%%%%%%%%%%%%%%%%%%

\section{Data}\label{sect:Data}
The main fitting results of our analysis are presented in Tables 1 and 2. An additional Table 3 has also been introduced to illustrate the correct normalization of our results with other existing studies in the literature when we use comparable data. This issue will be further discussed throughout our presentation. To generate the fitting results displayed in Tables 1,  2 and 3 we have run the MCMC code MontePython, together with CLASS, over an updated data set SNIa+$H(z)$+BAO+LSS+CMB, consisting of: i) 6 effective points on the normalized Hubble rate (including the covariance matrix) from the Pantheon+MCT sample \cite{Scolnic2018,Riess2018b}, which includes 1063 SNIa. As it is explained in \cite{Riess2018b}, the compression effectiveness of the information contained in such SNIa sample is extremely good; ii) 31 data from $H(z)$ from cosmic chronometers \cite{Jimenez2003,Simon,Stern,MorescoA,Zhang,MorescoB,MorescoC,Ratsismbazafy2017}; iii) 16 effective BAO points \cite{Kazin2014,GilMarin2016,Bourboux2017,Carter2018,GilMarin2018}; iv) 19 effective points from LSS, specifically 18 points from the observable $f(z)\sigma_8(z)$  (obtained from redshift-space distortions -- RSD)\,\cite{Guzzo09,Song09,Blake2011fs8,Beutler2012,Blake2013fs8,Simpson,Okumura2016,GilMarin2016,Howlett2017,Feng2018,GilMarin2018,Mohammad2018} and one effective point from the weak lensing observable $S_8\equiv \sigma_8(0)\left({\Omega_m}/{0.3}\right)^{0.5}$ \cite{Hildebrandt2017}; v) Finally we make use of the full CMB likelihood from Planck 2015 TT+lowP+lensing\,\cite{Planck}. It is important to emphasize that we take into account all the known correlations among data.  Owing to their special significance in this kind of analysis, we have collected the set of BAO and LSS  data points used in this work in the Tables 4 and 5, respectively.

The total data set just described will be referred to as DS1.  This is our baseline scenario since it involves a large sample of all sorts of cosmologial data and will be used to compare with the possible additional effects that emerge when we enrich its structure, as we shall comment in a moment. For a more detailed discussion of the data involved in DS1, see \cite{PLB2017,ApJL2015,MNRAS2018}.
In this study, however, we wish to isolate also the effect from the triad of  BAO+LSS+CMB data. These ingredients may be particularly sensitive to the DDE, as shown in the previous references. Such subset of DS1 will be called  DS2 and contains the same data as DS1 except SNIa+$H(z)$. In the next section we define further specifications of these two datasets with special properties.

\begin{table}[t]
\begin{center}
\resizebox{10cm}{!}{
%\begin{scriptsize}
\begin{tabular}{| c | c |c | c |c|c|}
\multicolumn{1}{c}{Survey} &  \multicolumn{1}{c}{$z$} &  \multicolumn{1}{c}{Observable} &\multicolumn{1}{c}{Measurement} & \multicolumn{1}{c}{{\small References}} & \multicolumn{1}{c}{{\small Data set}}
\\\hline
6dFGS+SDSS MGS & $0.122$ & $D_V(r_d/r_{d,fid})$[Mpc] & $539\pm17$[Mpc] &\cite{Carter2018} & SP/BSP
\\\hline

 WiggleZ & $0.44$ & $D_V(r_d/r_{d,fid})$[Mpc] & $1716.4\pm 83.1$[Mpc] &\cite{Kazin2014} & SP/BSP \tabularnewline
\cline{2-4} & $0.60$ & $D_V(r_d/r_{d,fid})$[Mpc] & $2220.8\pm 100.6$[Mpc]& &\tabularnewline
\cline{2-4} & $0.73$ & $D_V(r_d/r_{d,fid})$[Mpc] &$2516.1\pm 86.1$[Mpc] & &
\\\hline

DR12 BOSS (BSP) & $0.32$ & $Hr_d/(10^{3}km/s)$ & $11.549\pm0.385$   &\cite{GilMarin2016} & BSP \\ \cline{3-4}
 &  & $D_A/r_d$ & $6.5986\pm0.1337$ & &\tabularnewline \cline{3-4}
 \cline{2-2}& $0.57$ & $Hr_d/(10^{3}km/s)$  & $14.021\pm0.225$ & &\\ \cline{3-4}
 &  & $D_A/r_d$ & $9.3869\pm0.1030$ & &\\\hline

DR12 BOSS (SP) & $0.38$ & $D_M(r_d/r_{d,fid})$[Mpc] & $1518\pm22$   &\cite{Alam2017} & SP \\ \cline{3-4}
 &  & $H(r_{d,fid}/r_d)$[km/s/Mpc] & $81.5\pm1.9$ & &\tabularnewline \cline{3-4}
 \cline{2-2}& $0.51$ & $D_M(r_d/r_{d,fid})$[Mpc] & $1977\pm27$ & &\\ \cline{3-4}
 &  & $H(r_{d,fid}/r_d)$[km/s/Mpc] & $90.4\pm1.9$ & &\\ \cline{3-4}
 \cline{2-2}& $0.61$ & $D_M(r_d/r_{d,fid})$[Mpc]  & $2283\pm32$ & &\\ \cline{3-4}
 &  & $H(r_{d,fid}/r_d)$[km/s/Mpc] & $97.3\pm2.1$ & &\\\hline

eBOSS & $1.19$ & $Hr_d/(10^{3}km/s)$ & $19.6782\pm1.5866$   &\cite{GilMarin2018}& SP/BSP \\ \cline{3-4}
 &  & $D_A/r_d$ & $12.6621\pm0.9876$ & &\tabularnewline \cline{3-4}
 \cline{2-2}& $1.50$ & $Hr_d/(10^{3}km/s)$  & $19.8637\pm2.7187$ & &\\ \cline{3-4}
 &  & $D_A/r_d$ & $12.4349\pm1.0429$ & &\\ \cline{3-4}
 \cline{2-2}& $1.83$ & $Hr_d/(10^{3}km/s)$  & $26.7928\pm3.5632$ & &\\ \cline{3-4}
 &  & $D_A/r_d$ & $13.1305\pm1.0465$ & &\\\hline

Ly$\alpha$-forest & $2.40$ & $D_H/r_d$ & $8.94\pm0.22$   &\cite{Bourboux2017} & SP/BSP
\\ \cline{3-4} &  & $D_M/r_d$ & $36.6\pm1.2$ & &\\\hline

\end{tabular}}
% \end{scriptsize}
\caption{\scriptsize Published values of BAO data {used in the main analyses. In the last column we specify the data sets in which these points have been employed. We label the latter with SP if they are used in DS1/SP; with BSP if they are used in DS1/BSP and DS2/BSP; and finally with SP/BSP if they are used in the three data sets, DS1/SP, DS1/BSP, and DS2/BSP. See the quoted references, and the text in Sect. 3.}}
\end{center}
\end{table}
%%%%%%%%%%%%%%%%%%%%%%%%%%%%%%%%%%%%%%%%%%%%%%%%%%%%%%%%%%%%%%%%%%%%%%%%%%%%%%
%

%
%%%%%%%%%%%%%%%%%%%%%%%%%%%%%%%%%%%%%%%%%%%%%%%%%%%%%%%%%%%%%%%%%%%%%%%%%%%%%%
%
\begin{table}[t]
\begin{center}
\resizebox{7cm}{!}{
%\begin{scriptsize}
\begin{tabular}{| c | c |c | c |c |}
\multicolumn{1}{c}{Survey} &  \multicolumn{1}{c}{$z$} &  \multicolumn{1}{c}{$f(z)\sigma_8(z)$} & \multicolumn{1}{c}{{\small References}} & \multicolumn{1}{c}{{\small Data set}}
\\\hline
2MTF & $0$ & $0.505\pm 0.084$ & \cite{Howlett2017} & SP/BSP
\\\hline
6dFGS & $0.067$ & $0.423\pm 0.055$ & \cite{Beutler2012} & SP/BSP
\\\hline
SDSS-DR7 & $0.10$ & $0.376\pm 0.038$ & \cite{Feng2018} & SP/BSP
\\\hline
GAMA & $0.18$ & $0.29\pm 0.10$ & \cite{Simpson} & SP/BSP
\\ \cline{2-4}& $0.38$ & $0.44\pm0.06$ & \cite{Blake2013fs8} &
\\\hline
DR12 BOSS (BSP) & $0.32$ & $0.427\pm 0.056$  & \cite{GilMarin2016} &  BSP\\ \cline{2-3}
 & $0.57$ & $0.426\pm 0.029$ & & \\\hline
DR12 BOSS (SP) & $0.38$ & $0.497\pm 0.045$ & \cite{Alam2017}&  SP\tabularnewline
\cline{2-3} & $0.51$ & $0.458\pm0.038$ & &\tabularnewline
\cline{2-3} & $0.61$ & $0.436\pm0.034$ & &

\\\hline
 WiggleZ & $0.22$ & $0.42\pm 0.07$ & \cite{Blake2011fs8} & SP/BSP\tabularnewline
\cline{2-3} & $0.41$ & $0.45\pm0.04$ & &\tabularnewline
\cline{2-3} & $0.60$ & $0.43\pm0.04$ & &\tabularnewline
\cline{2-3} & $0.78$ & $0.38\pm0.04$ & &
\\\hline
VIPERS & $0.60$ & $0.49\pm 0.12$ & \cite{Mohammad2018}& SP/BSP
\\ \cline{2-3}& $0.86$ & $0.46\pm0.09$ & &
\\\hline
VVDS & $0.77$ & $0.49\pm0.18$ & \cite{Guzzo09},\cite{Song09}& SP/BSP
\\\hline
FastSound & $1.36$ & $0.482\pm0.116$ & \cite{Okumura2016}& SP/BSP
\\\hline
eBOSS & $1.19$ & $0.4736\pm 0.0992$ & \cite{GilMarin2018}& SP/BSP \tabularnewline
\cline{2-3} & $1.50$ & $0.3436\pm0.1104$ & &\tabularnewline
\cline{2-3} & $1.83$ & $0.4998\pm0.1111$ & &

\\\hline
 \end{tabular}}
\caption{\scriptsize{As in Table 4, but for the published values of $f(z)\sigma_8(z)$. See the quoted references, and text in Sect. 3.}}
\end{center}
\end{table}
%%%%%%%%%%%%%%%%%%%%%%%%%%%%%%%%%%%%%%%%%%%%%%%%%%%%%%%%%%%%%%%%%%%%%%%%%%%%%%
%

\section{Spectrum versus bispectrum}\label{sect:SpectrumBispectrum}

The usual analyses of structure formation data in the literature are performed in terms of the matter power spectrum  $P({\bf k})$, referred to here simply as spectrum (SP). As we know, the latter is defined in terms of the two-point correlator of the density field $D({\bf k})$ in Fourier space, namely $\langle D({\bf k})\,D({\bf k}')\rangle=\delta({\bf k}+{\bf k}') P({\bf k})$, in which $\delta$ is the Dirac delta of momenta. For a purely Gaussian distribution, any higher order correlator of even order decomposes into sums of products of two-point functions, in a manner very similar to Wick's theorem in QFT. At the same time, all correlators of odd order vanish. This ceases to be true for non-Gaussian distributions, and the first nonvanishing correlator is then the bispectrum $B({\bf k}_1,{\bf k}_2,{\bf k}_3)$, which is formally connected to the three-point function
\begin{equation}\label{eq:bispectrum}
\langle D({\bf k}_1)\,D({\bf k}_2)\, D({\bf k}_3)\rangle= \delta({\bf k}_1+{\bf k}_2+{\bf k}_3)B({\bf k}_1,{\bf k}_2,{\bf k}_3)\,.
\end{equation}
The Dirac $\delta$ selects in this case all the triangular configurations. Let us note that even if the primeval spectrum would be purely Gaussian, gravity makes fluctuations evolve non-Gaussian. Such deviations with respect to a normal distribution may be due both to the evolution of gravitational instabilities that are amplified from the initial perturbations, or even from some intrinsic non-Gaussianity of the primordial spectrum.  For example, certain implementations of inflation (typically multifield inflation models) unavoidably lead to a certain degree of non-Gaussianity\,\cite{Liddle}.  Therefore,  in practice  the bispectrum is expected to be a nonzero parameter in real cosmology, even if starting from perfect Gaussianity, which is in no way an absolute condition to be preserved.  On the other hand, such departure  should, of course,  be small.  But the dynamics of the DE is also expected to be small, so  there is a possible naturalness relationship between the two.

The bispectrum (BSP) has been described in many places in the literature, see e.g.\,\cite{Liddle,DEBookAmendola}  and references therein.  The physical importance of including the bispectrum cannot be overemphasized as it furnishes important complementary information that goes beyond the spectrum. If fluctuations in the structure formation were strictly Gaussian, their full statistical description would be contained in the two-point correlation function  $\langle D({\bf k})\,D({\bf k}')\rangle$ since, as already mentioned,  all  higher order correlators of even order can be expanded in terms of products of two-point functions.  In such a case the formal bispectrum defined above would identically vanish.  However,  there is no a priori reason for that to happen, and in general this is not what we expect if we take into consideration the reasons mentioned above. The crucial question is:  how to test the real situation in practice?  While the above definitions are the formal ones, operationally (in other words, at the practical level of galaxy counting) one must resort to use SP and BSP estimators of empirical nature.  For the power spectrum estimator one may use $\langle F_2({\bf k}_1)F_2({\bf k}_2)\rangle$, where $F_2({\bf q})$ is the Fourier transform of an appropriately defined weighted field of density fluctuations, that is to say, one formulated in terms of the number density of galaxies\,\cite{GilMarin2016}.

Similarly, a bispectrum estimator $\langle F_3({\bf k}_1)F_3({\bf k}_2) F_3({\bf k}_3)\rangle$ can be defined from the angle-average of closed triangles defined by the $\bf k$-modes, ${\bf k}_1,\,{\bf k}_2,\,{\bf k}_3$, where $F_3({\bf q})$ is the Fourier transform of the corresponding weighted field of density fluctuations defined  in terms of the number density of galaxies. It can be conveniently written as
\begin{equation}
 \langle F_3({\bf k}_1)F_3({\bf k}_2) F_3({\bf k}_3)\rangle=\frac{k_f^3}{V_{123}}\int d^3{\bf r}\, \mathcal{D}_{\mathcal{S}_1}({\bf r}) \mathcal{D}_{\mathcal{S}_2}({\bf r}) \mathcal{D}_{\mathcal{S}_3}({\bf r})\,,
 \label{eq:bis2}
\end{equation}
i.e. through an expression involving a separate product of Fourier integrals
\begin{equation}
 \mathcal{D}_{\mathcal{S}_j}({\bf r})\equiv \int_{\mathcal{S}_j} d{\bf q}_j\, F_3({\bf q}_j)e^{i{\bf q}_j\cdot{\bf r}}\,.
\end{equation}
Here $k_f$ is the fundamental frequency, $k_f=2\pi/L_{\rm box}$, $L_{\rm box}$ the size of the box in which the galaxies are embedded and
\begin{equation}
 V_{123}\equiv\int_{\mathcal{S}_1} d{\bf q}_1\, \int_{\mathcal{S}_2} d{\bf q}_2\, \int_{\mathcal{S}_3} d{\bf q}_3\, \delta({\bf q}_1+{\bf q}_2+{\bf q}_3)
\end{equation}
is the number of fundamental triangles inside the shell defined by $\mathcal{S}_1$, $\mathcal{S}_2$ and $\mathcal{S}_3$, with $\mathcal{S}_i$ the region of the $k$-modes contained in a $k$-bin, $\Delta k$, around $k_i$. The Dirac $\delta$ insures that only closed triangles are included -- see the mentioned references for more details. The measurement of the bispectrum estimator $\langle F_3({\bf k}_1)F_3({\bf k}_2) F_3({\bf k}_3)\rangle$  is essential to be sensitive to possible higher order effects associated to non-Gaussianities in the distribution of galaxies. This task is what has been done in the important work\,\cite{GilMarin2016}.

Here we wish to dwell on the impact of the bispectrum as a potential tracer of the DDE. Observationally, the data on BAO+ LSS (more specifically, the $f\sigma_8$ part of LSS) including both the spectrum (SP) and bispectrum (BSP) are taken from \cite{GilMarin2016}, together with the correlations among these data encoded in the provided covariance matrices. The same data  including SP but no BSP has been considered in \cite{Alam2017}. In this study, we analyze the full dataset DS1 with spectrum only (dubbed DS1/SP) and also the same data when we include both SP and BSP (denoted DS1/BSP for short). In addition, we test the DDE sensitivity of the special subset DS2, which involves both SP+BSP components (scenario DS2/BSP). The contrast of results between the ``bispectrumless'' scenarios (i.e. the pure SP ones) and those including the matter bispectrum component as well (i.e. the SP+BSP ones)   will be made apparent in our study, specially through our devoted discussions in sections 5 and 6.

%%%%%%%%%%%%%%%%%%%%%%%%%%%%%%%%%%%%%%%%%%%%%%%%%%%%%%%%%%%%%%
%%%%
\begin{figure}[!t]
\begin{center}
\label{LCDMtriangular}
\includegraphics[width=5.5in, height=4.7in]{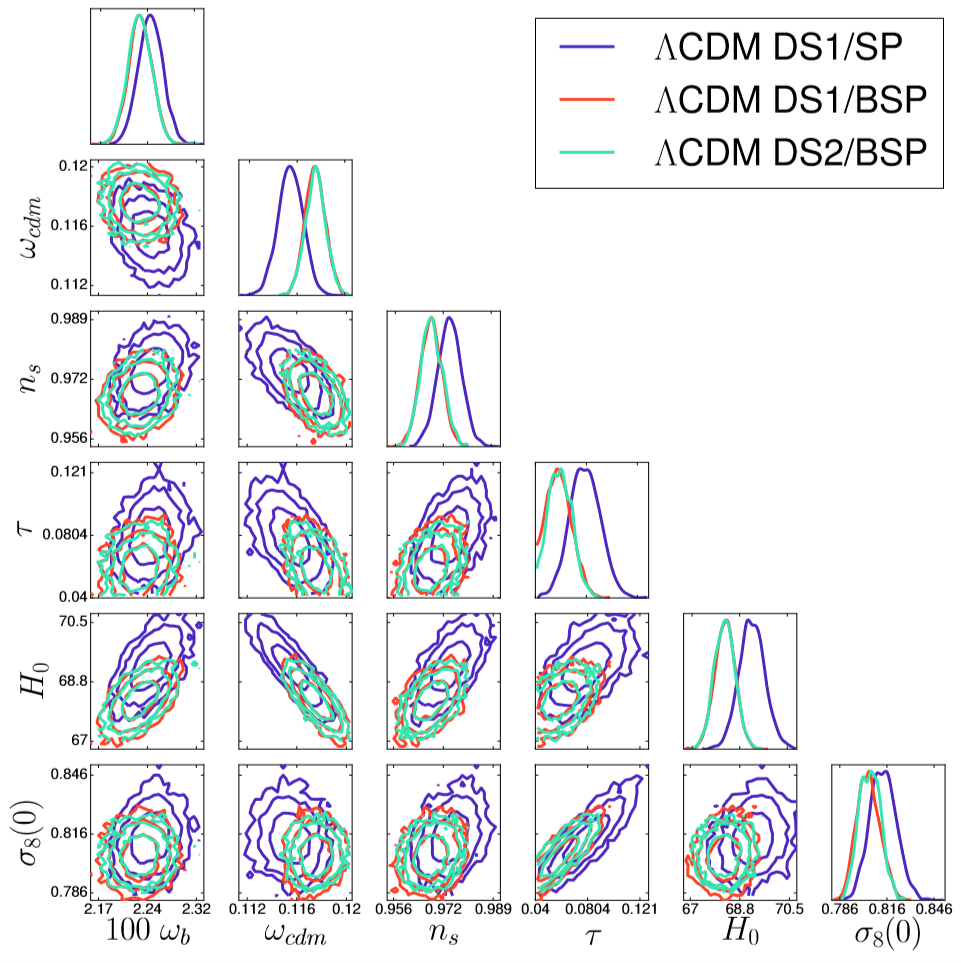}
\caption{\scriptsize Triangular matrix containing all the possible combinations of two-dimensional marginalized distributions for the various $\Lambda$CDM fitting parameters (at $1\sigma$, $2\sigma$ and $3\sigma$ c.l.), together with the corresponding one-dimensional marginalized likelihoods for each parameter. $H_0$ is expressed in km/s/Mpc. We present the results for the main data sets used in this work: DS1/SP, DS1/BSP, and DS2/BSP (cf. Tables 1 and 2 for the numerical fitting results, and the text for further details).}
\end{center}
\end{figure}
%%%%%%
%%%%%%%%%%%%%%%%%%%%%%%%%%%%%%%%%%%%%%%%%%%%%%%%%%%%%%%%%%%%%%%%%
%%%%%%%%%%%%%%%%%%%%%%%%%%%%%%%%%%%%%%%%%%%%%%%%%%%%%%%%%%%%%%%%%

The numerical fitting results that we have obtained for the DDE models under examination  and the $\CC$CDM are displayed  in Tables 1 and 2. While we use Planck 2015 CMB data with full likelihood throughout most of our analysis, in the second block of Table 2 we report on the preliminary results obtained from the recent Planck 2018 CMB data under compressed likelihood  \cite{Planck2018,Chen2018}.  Let us note that the full likelihood for  Planck 2018 CMB data is not public yet.   We discuss the  results that we have obtained and their possible implications in the next two sections.

As indicated above, an additional fitting table has  been generated (Table 3) in order to illustrate the fact that the results presented in our work are consistent with other studies presented in the literature, namely for the case when we restrict our analysis to the baseline bispectrumless scenario introduced in our work (i.e.  DS1/SP).  We substantiate this fact by introducing an alternative bispectrumless  scenario, which we call  DS3/SP.  The latter is essentially coincident with the one used in the  investigations of the same models recently presented by the authors of \cite{ParkRatraXCDM,ParkRatraPhiCDM}, although there exist some mild differences between our DS3/SP scenario and the one considered by these authors, to wit: (i) In the supernovae sector there appear two differences:  first, we use the compressed Pantheon+MCT data\,\cite{Scolnic2018}  mentioned in the previous section, which in expanded form includes more than one thousand supernovae,  whereas they use the full (uncompressed) list of supernova data from the Pantheon compilation, but in their  list -- and here lies the second difference--  they do not include the 15 high-redshift SNIa ($z>1$) from the CANDELS and CLASH Multy-Cycle Treasury (MCT) programs. However these differences should not be significant at all, and we refer the reader to the detailed discussion in Ref.\,\cite{Riess2018b} to justify why is so. In fact, in the latter  reference it is demonstrated  that the constraints derived from the full Pantheon+MCT compilation of supernovae of Type Ia are essentially indistinguishable from those derived from the corresponding compressed likelihood (see Fig. 3 of that reference); (ii)  Concerning the BOSS Ly$\alpha$-forest data, they use the exact distributions, whereas we use the Gaussian approximations. A direct comparison of the values of the fitting parameters reported in \cite{ParkRatraXCDM,ParkRatraPhiCDM} with those presented in our Table 3 confirms a high degree of compatibility, being the discrepancies in all cases  a small fraction of  $ 1\sigma$. For instance, for the XCDM we obtain $w_0=-1.002^{+0.034}_{-0.035}$, and the authors of \cite{ParkRatraXCDM} find $w_0=-0.994\pm 0.033$, which differs by less than  $0.17\sigma$  from the former, taking the errors in quadrature.  The difference with respect to the  value of $w_0$  corresponding to the DS1/SP dataset in Table 1 is also small:  $0.18\sigma$. On these grounds we judge that the approach adopted in our analysis is fully consistent and  that the tiny differences which may appear between the exact and compressed treatment of the SNIa and the exact and Gaussian Ly$\alpha$-forest likelihoods cannot be held responsible for any significant change in the main conclusions of our work as to the dynamical nature of the DE. Put another way, the preliminary signal of DDE that we find in this work cannot be attributed to the tiny differences among the bispectrumless  data sets existing in the literature, whether our DS1/SP, DS3/SP,  the exact one used e.g. by \cite{ParkRatraXCDM,ParkRatraPhiCDM} or any other consistent dataset employed by different authors.  All these bispectrumless scenarios are statistically equivalent since they involve a sufficiently complete and uncorrelated set of data from the various SNIa+$H(z)$+BAO+LSS+CMB sources  and therefore they are mutually consistent within a fraction of $ 1\sigma$ errors.  We must conclude that the emerging indications of DDE that we have encountered should rather originate from the peculiarities of the bispectrum component, which seems to be particularly sensitive to the dynamical features presumably sitting in the DE.

%%%%%%%%%%%%%%%%%%%%%%%%%%%%%%%%%%%%%%%%%%%%%%%%%%%%%%%%%%%%%%
%%%%
\begin{figure}
\begin{center}
\label{XCDMtriangular}
\includegraphics[width=5.5in, height=4.7in]{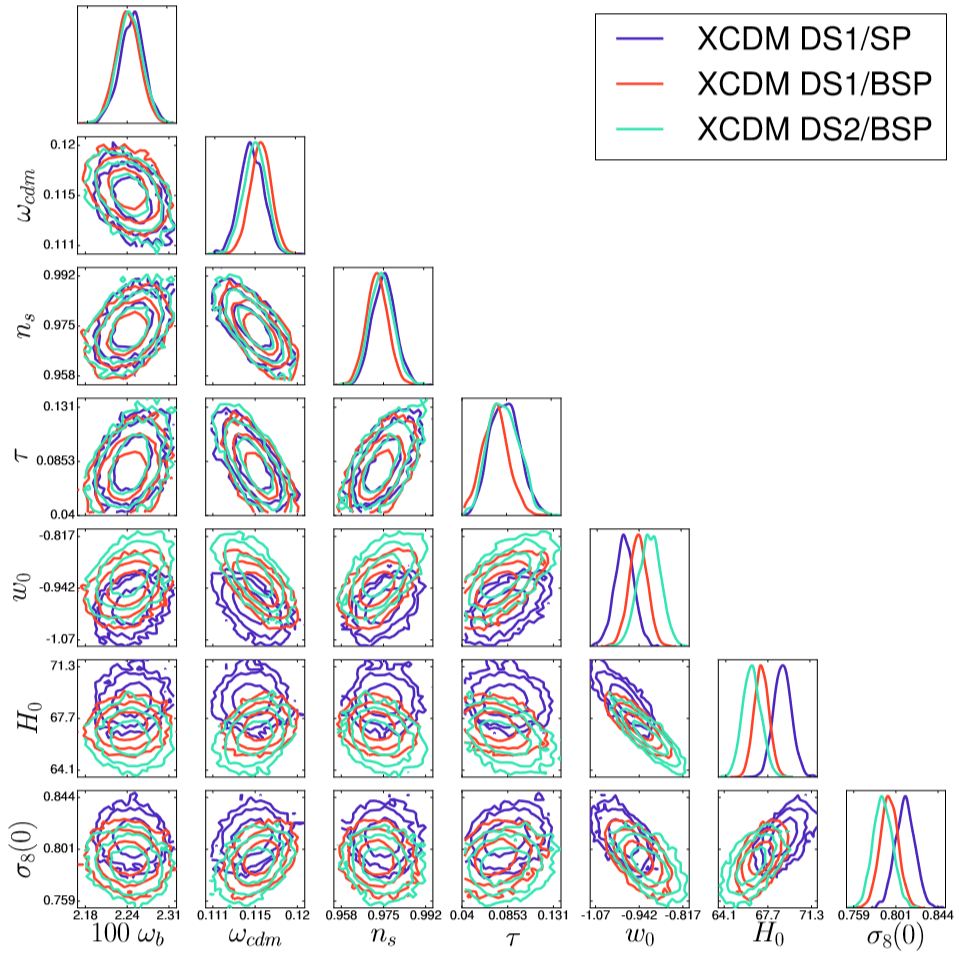}
\caption{\scriptsize  As in Fig. 3, but for the XCDM. Here we also show the two-dimensional contours and one-dimensional distribution for the EoS parameter $w_0$.}
\end{center}
\end{figure}
%%%%%%
%%%%%%%%%%%%%%%%%%%%%%%%%%%%%%%%%%%%%%%%%%%%%%%%%%%%%%%%%%%%%%%%%
%%%%%%%%%%%%%%%%%%%%%%%%%%%%%%%%%%%%%%%%%%%%%%%%%%%%%%%%%%%%%%%%%

\section{Confronting DDE to observations}\label{sect:LinStructure}

As we can see from Tables 1 and 2, the comparison of the $\CC$CDM with the DDE models points to a different sensibility of the data sets used to the dynamics of the DE.   If we start focusing on the results from the  bispectrumless scenario DS1/SP -- cf.  left block of Table 1 -- there is no evidence that the DDE models perform better than the $\CC$CDM. The XCDM, for instance,  yields a weak signal which is compatible with $w_0=-1$ (i.e. a rigid CC). This is consistent e.g. with the analysis of  \cite{ParkRatraXCDM}.  The $\phi$CDM model remains also inconclusive under the same data. The upper bound of  $\alpha<0.092$ at $1\sigma$  ($0.178$ at 2$\sigma$) recorded in that table  is consistent, too,  with the recent studies of\,\cite{ParkRatraPhiCDM}.  The same conclusion is  derived upon inspection of Table 3, based on the alternative bispectrumless  scenario DS3/SP, whose fitting results are indeed  consistent with those collected for our DS1/SP as well as with those from the mentioned references \cite{ParkRatraXCDM} and \cite{ParkRatraPhiCDM}.

%%%%%%%%%%%%%%%%%%%%%%%%%%%%%%%%%%%%%%%%%%%%%%%%%%%%%%%%%%%%%%
%%%%
\begin{figure}
\begin{center}
\label{phiCDMtriangular}
\includegraphics[width=5.5in, height=4.7in]{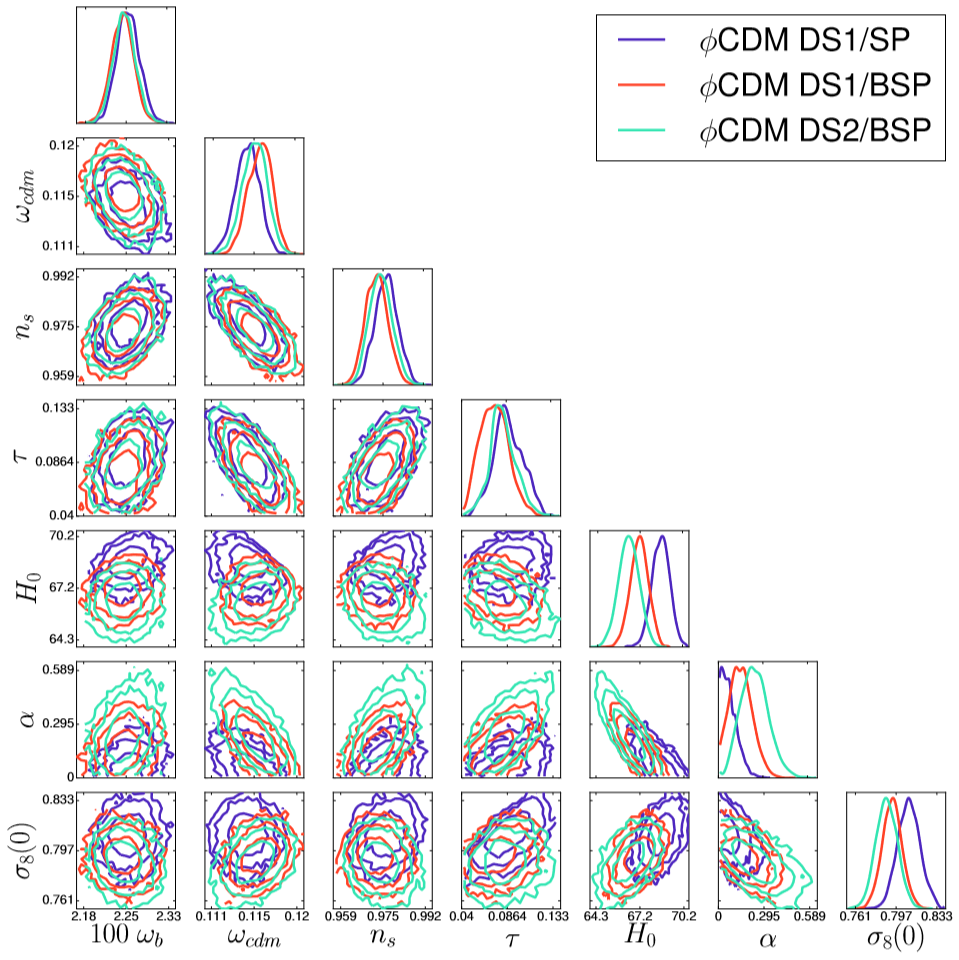}
\caption{\scriptsize As in Fig. 3, but for the $\phi$CDM model. We also include the two-dimensional contours and one-dimensional distribution for the parameter $\alpha$ of the potential (\ref{eq:PRpotential}).}
\end{center}
\end{figure}
%%%%%%
%%%%%%%%%%%%%%%%%%%%%%%%%%%%%%%%%%%%%%%%%%%%%%%%%%%%%%%%%%%%%%%%%
%%%%%%%%%%%%%%%%%%%%%%%%%%%%%%%%%%%%%%%%%%%%%%%%%%%%%%%%%%%%%%%%%

Notwithstanding, both the XCDM parametrization and the $\phi$CDM model fare significantly better than the $\CC$CDM
 if we consider the dataset DS1/BSP, i.e. upon including the bispectrum component of the BAO+LSS data.  The corresponding DDE signature is  about $2\sigma$ c.l. However, it is further enhanced within the restricted DS2/BSP dataset, where the XCDM and the $\phi$CDM reach in between  $2.5-3\sigma$ c.l. (cf. left  block of Table 2).  As for the CPL parametrization, Eq.\, (\ref{eq:CPL}), we  record it explicitly in Table 1 for the DS1/BSP case only.  One can check that even in this case the errors in the EoS parameters are still too big to capture any clear sign of DE dynamics, owing to the additional parameter present in this model. Specifically, we find $w_0= -0.934^{+0.067}_{-0.075}$ and $w_1 = -0.045 ^{+0.273}_{-0.204}$,  hence fully compatible with a rigid CC ($w_0=-1, w_1=0$).
 %We shall not consider the CPL any longer in our considerations, since it is disfavored by the Occam razor criterion.}

%%%%%%%%%%%%%%%%%%%%%%%%%%%%%%%%%%%%%%%%%%%%%%%%%%%%%%%%%%%%%%%%%
%%%%%%%%%%%%%%%%%%%%%%%%%%%%%%%%%%%%%%%%%%%%%%%%%%%%%%%%%%%%%%%%%

In Fig. 1 we show the contour plots  in the $(\omega_{cdm},w_0)$ and $(\omega_{cdm},\alpha)$ planes for the XCDM and $\phi$CDM models, respectively,  at different confidence levels, which are obtained with the (bispectrumless) DS1/SP data set.  It is obvious from these contours that the models do not exhibit a clear preference for  DDE.  The contour plots for the XCDM (on the left side of Fig. 1)  appear located roughly $50\%$ up and  $50\%$  down with respect to the cosmological constant divide $w_0=-1$, and therefore we cannot appraise any marked preference for a deviation into the quintessence ($w_0\gtrsim-1$) or the phantom region ($w_0\lesssim-1$). Similarly,  the contour plots for the parameter $\alpha$ of the  $\phi$CDM model (indicated on the right side of Fig. 1)  show that $\alpha$ is consistent with zero at $1\sigma$, which means that the potential  (\ref{eq:PRpotential}) is perfectly compatible with a cosmological constant.

%%%%%%%%%%%%%%%%%%%%%%%%%%%%%%%%%%%%%%%%%%%%%%%%%%%%%%%%%%%%%%%%%
%%%%%%%%%%%%%% FIGURE 2 %%%%%%%%%%%%%%%%%%%%%%%%%%%
%%%%%%%%%%%%%%%%%%%%%%%
%%%%
\begin{figure}
\begin{center}
\label{XCDMPlanck2018}
\includegraphics[width=4.6in, height=2.7in]{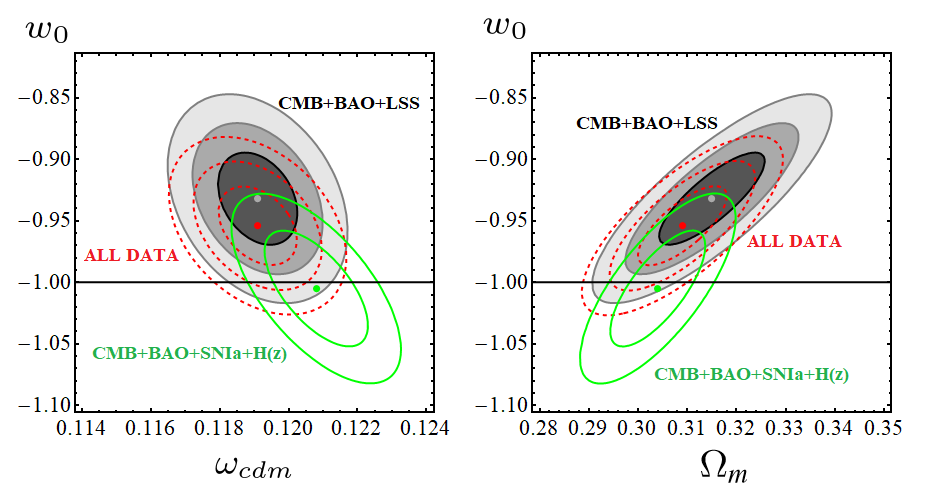}
\caption{\scriptsize Contour lines for the XCDM in the planes $(\omega_{cdm},w_0)$ and $(\Omega_m,w_0)$ obtained with different data set combinations and using the compressed Planck 2018 data \cite{Planck2018,Chen2018} instead of the full Planck 2015 likelihood \cite{Planck}. More concretely we show the results obtained with the DS2/BSP data set (CMB+BAO+LSS, in grayish shades, see the second block of Table 2), the DS1/BSP (ALL DATA, dashed red contours), and the DS1/BSP but without LSS (CMB+BAO+SNIa+$H(z)$, solid green contours). The solid points indicate the location of the best-fit values in each case. When the LSS data points with bispectrum are considered in combination with the CMB and BAO constraints the DDE signal exceeds the $2\sigma$ c.l., regardless we use the Planck 2015 or 2018 information. See the text for further details.}
\end{center}
\end{figure}
%%%%%%
%%%%%%%%%%%%%%%%%%%%%%%%%%%%%%%%%%%%%%%%%%%%%%%%%%%%%%%%%%%%%%%%%
%%%%%%%%%%%%%%%%%%%%%%%%%%%%%%%%%%%%%%%%%%%%%%%%%%%%%%%%%%%%%%%%%

In Fig. 2, however,  the situation has changed in a rather conspicuous way.  Once more  we display the corresponding  contour plots for the XCDM and $\phi$CDM models, but now they are obtained in the presence of the bispectrum, specifically we use in this case the DS2/BSP data set.  While we could use the entire  DS1/BSP set, the former subset (made exclusively  on BAO+LSS+CMB)  is slightly  more sensitive and we choose it to illustrate which data  tend to optimize the DDE signal.  After all not all data is expected to be equally sensitive, and this is something that has already been noted for other DDE models\,\cite{MNRAS2018}.  Coming back to Fig. 2 we can see that, in stark contrast with Fig. 1, there is a marked preference for the contours of the XCDM to shift upwards into the quintessence region.  Specifically,  the EoS parameter $w_0$  lies now more than $2.5\sigma$ away from $-1$  in the quintessence domain  $w_0\gtrsim-1$.  On the other hand, in the $\phi$CDM case (shown in the right plot of the same figure)  we consistently find $\alpha>0$ at a similar (actually slightly higher) c.l.

%%%%%%%%%%%%%%%%%%%%%%%%%%%%%%%%%%%%%%%%%%%%%%%%%%%%%%%%%%%%%%
%%%%
\begin{figure}
\begin{center}
\label{phiCDMPlanck2018}
\includegraphics[width=4.6in, height=2.7in]{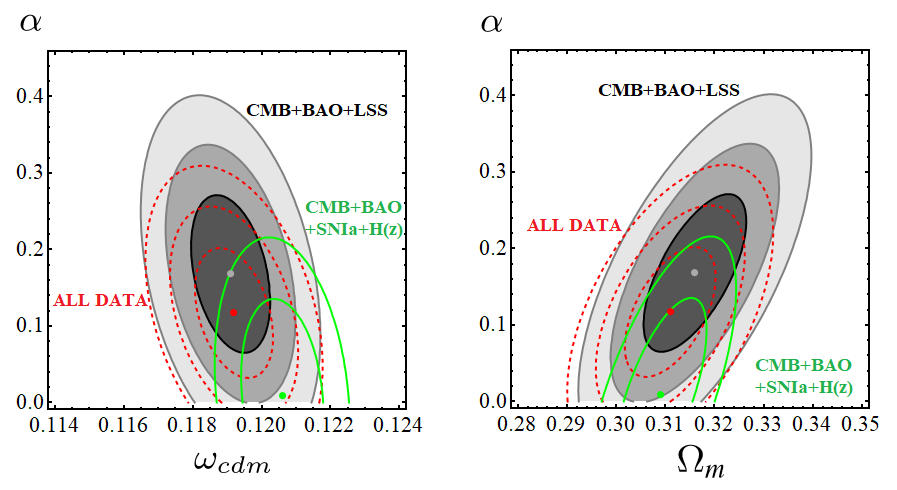}
\caption{\scriptsize The same as in Fig. 6, but for the $\phi$CDM model and in the planes $(\omega_{cdm},\alpha)$ and $(\Omega_m,\alpha)$.}
\end{center}
\end{figure}
%%%%%%
%%%%%%%%%%%%%%%%%%%%%%%%%%%%%%%%%%%%%%%%%%%%%%%%%%%%%%%%%%%%%%%%%
%%%%%%%%%%%%%%%%%%%%%%%%%%%%%%%%%%%%%%%%%%%%%%%%%%%%%%%%%%%%%%%%%

Figures 3-5 are also very illustrative. They contain the confidence contours in all the relevant planes of the parameter space up to $3\sigma$ c.l., together with the marginalized one-dimensional posterior distributions for the various fitting parameters. In Fig. 3 we present the results obtained with the DS1/SP, DS1/BSP and DS2/BSP scenarios in the context of the $\Lambda$CDM. Figs. 4 and 5 contain the analogous plots for the XCDM and $\phi$CDM, respectively. These figures allow us to appreciate in a very straightforward way what are the shifts in the confidence regions caused by the various changes of our cosmological data sets. Fig. 3 shows, for instance, that in the $\Lambda$CDM the central values of all the parameters that enter the fit get shifted away from their DS1/SP values  when the bispectrum information is also taken into account. These shifts are specially important for the parameters $\omega_{cdm}$, $n_s$, $\tau$ and $H_0$, which in the DS1/BSP scenario are shifted $+1.48\sigma$, $-1.13\sigma$, $-1.23\sigma$ and $-1.62\sigma$ away with respect to the DS1/SP case. In contrast, $\omega_b$ and $\sigma_8(0)$ are lesser modified ($-0.60\sigma$ and $-0.79\sigma$, respectively). The removal of the SNIa+$H(z)$ data points from the DS1/BSP data set (which gives rise to the DS2/BSP one) does not produce any additional significant change in the $\Lambda$CDM, as it can be easily checked by comparing the DS1/BSP and DS2/BSP contours and the corresponding one-dimensional distributions of Fig. 3. In the DDE models the effect on the parameters introduced by the bispectrum signal is quite different. To begin with, the values of $\omega_{cdm}$, $\omega_b$, $n_s$ and $\tau$ remain very stable and completely compatible at $<1\sigma$ regardless of the data set under consideration. Moreover, the values of these parameters are also compatible with those of the $\Lambda$CDM in the DS1/SP scenario. This means that the bispectrum does not force in the XCDM and $\phi$CDM any important shift of the central values of $\omega_{cdm}$, $\omega_b$, $n_s$ and $\tau$ with respect to the typical values found in the $\Lambda$CDM when we only consider the matter power spectrum part of the LSS+BAO data sector. Conversely, when we move from the DS1/SP scenario to the DS1/BSP and, finally, to the DS2/BSP, we find a progressive and non-negligible displacement of the peaks of the one-dimensional distributions for $H_0$ and $\sigma_8(0)$ towards lower values, which is accompanied by a departure of the DDE parameters of the XCDM ($w_0$) and $\phi$CDM ($\alpha$) from the $\Lambda$CDM values ($-1$ and $0$, respectively). The global decrease of $\sigma_8(0)$ from DS1/SP to DS2/BSP  is  $ -1.6\sigma$ ($-1.3\sigma$) for XCDM ($\phi$CDM), and the one of $H_0$ is  $-2.3\sigma$ for both models. This fact allows to reduce the well-known $\sigma_8$-tension \cite{Macaulay2013} in a more efficient way in the context of the DDE models. However, the low values of the Hubble parameter derived in the XCDM and $\phi$CDM from our fitting analyses keep the statistical tension with the local distance ladder determinations of \cite{Riess2016,Riess2018} high, reaching the latter the $\sim 3.8\sigma$ level. Nevertheless, it is important to remark that the values of $H_0$ obtained by us in the current study are  perfectly consistent with those obtained by some authors that apply model-independent reconstruction techniques and low-redshift data from SNIa, BAO and $H(z)$ data points from cosmic chronometers, see e.g. \cite{YuRatraWang,H0GomezValentAmendola,H0Feeney,H0Haridasu,H0DES}. In the DDE scenarios analyzed in this paper, the $\sigma_8$-tension is directly linked with the $H_0$-tension. We can only loosen the former at the expense of keeping the latter (cf. the contour lines in the $H_0$-$\sigma_8(0)$ plane of Figs. 4 and 5). This is consistent with our analysis of \cite{PLB2017}, in which we  studied other DDE models pointing to the same conclusion. Future data might be able to elucidate the ultimate origin of the $H_0$-tension. If the true value of $H_0$ lies in the Planck-preferred region of $H_0\sim 66-68$ km/s/Mpc, then the DDE models can offer an efficient way of automatically relieving the $\sigma_8$-tension.

%%%%%%%% FIGURE 4 %%%%%%%%%%%%%%%%%%%%%%%%%%%%%%%%%
\begin{figure}
\begin{center}
\label{FigLSS2}
\includegraphics[width=4.5in, height=3in]{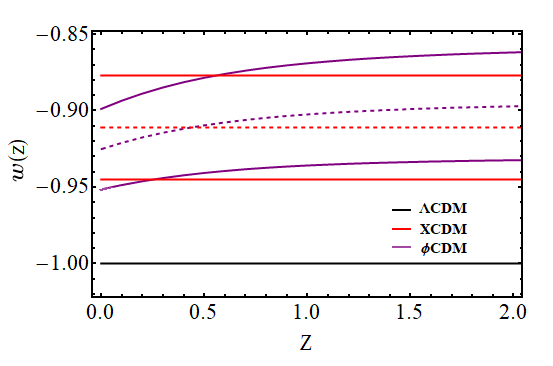}
\caption{\scriptsize  The EoS for the considered DE models
within the corresponding $1\sigma$ bands and under scenario DS2/BSP. For the XCDM the EoS
is constant and points to quintessence at  $\sim 2.6\sigma$ c.l (cf. Table 2). For the $\phi$CDM the EoS evolves with time and
is computed through a Monte Carlo analysis. The
current value is $w(z=0) = - 0.925\pm 0.026$ and favors once more
quintessence at  $\sim 2.9\sigma$ c.l.}
\end{center}
\end{figure}
%\addtolength{\dbltextfloatsep}{-0.5cm}
%%%%%%%%%%%%%%%%%%%%%%%%%%%%%%%%%%%%%%%%%%%%%%%%%

From a more prospective point of view, in Figs. 6 and 7 we perform a preliminary exploration of  the corresponding results using  Planck 2018 data. The results for the same DDE models are obtained under different data set combinations, but on this occasion making use of the Planck 2018 compressed CMB data, given the mentioned fact that the full likelihood for Planck 2018 is not publicly available yet. These plots make, once more, palpable the importance of the bispectrum  for the study of the DDE.  Indeed,  it is only when the bispectrum is included in the analysis, together with the CMB and BAO data sets,  that a non-negligible signal in favor of the dynamical nature of the DE clearly pops out.

In Fig. 8 we plot the EoS of the various models in terms of the redshift near our time and  within the $1\sigma$ error bands for the XCDM and $\phi$CDM,  again for the DS2/BSP case. These bands  have been computed from a Monte Carlo  sampling of the $w(z)$ distributions.  The behavior of the curves shows that the quintessence-like behavior is sustained until the present epoch. For the $\phi$CDM we find  $w(z=0) = - 0.925\pm 0.026$, thus implying a DDE signal at  $\sim 2.9\sigma$ c.l., which is consistent with the XCDM result\footnote{We point out that the higher c.l. reported in \cite{MPLA2017} for these models originated from the unexpected change in the reported errors of the LSS data  between the original version of \cite{{GilMarin2016}} (i.e. their version v1)  as compared to the published version (v2), in which the errors are slightly bigger.}.

Finally, in Fig. 9 we compute the matter power spectrum and the temperature anisotropies for the DDE models using Planck 2015 data, and also display the percentage differences with respect to the $\CC$CDM. As can be seen, the differences of the CMB anisotropies between the DDE models and the $\CC$CDM remain safely small,  of order $\sim1\%$ at most for the entire range.  The $3-5\%$ suppression of the matter power spectrum $P(k)$ of the DDE models in the relevant range of wave numbers $k$ with respect to the one of the concordance model is what gives rise to lower values of $\sigma_8(0)$ (cf. Tables 1-3).

%%%%%%     FIGURE 5  %%%%%%%%%%%%%%%%%%%%%%%%
\begin{figure}
\begin{center}
\label{FigLSS3}
\includegraphics[width=6in, height=3.5in]{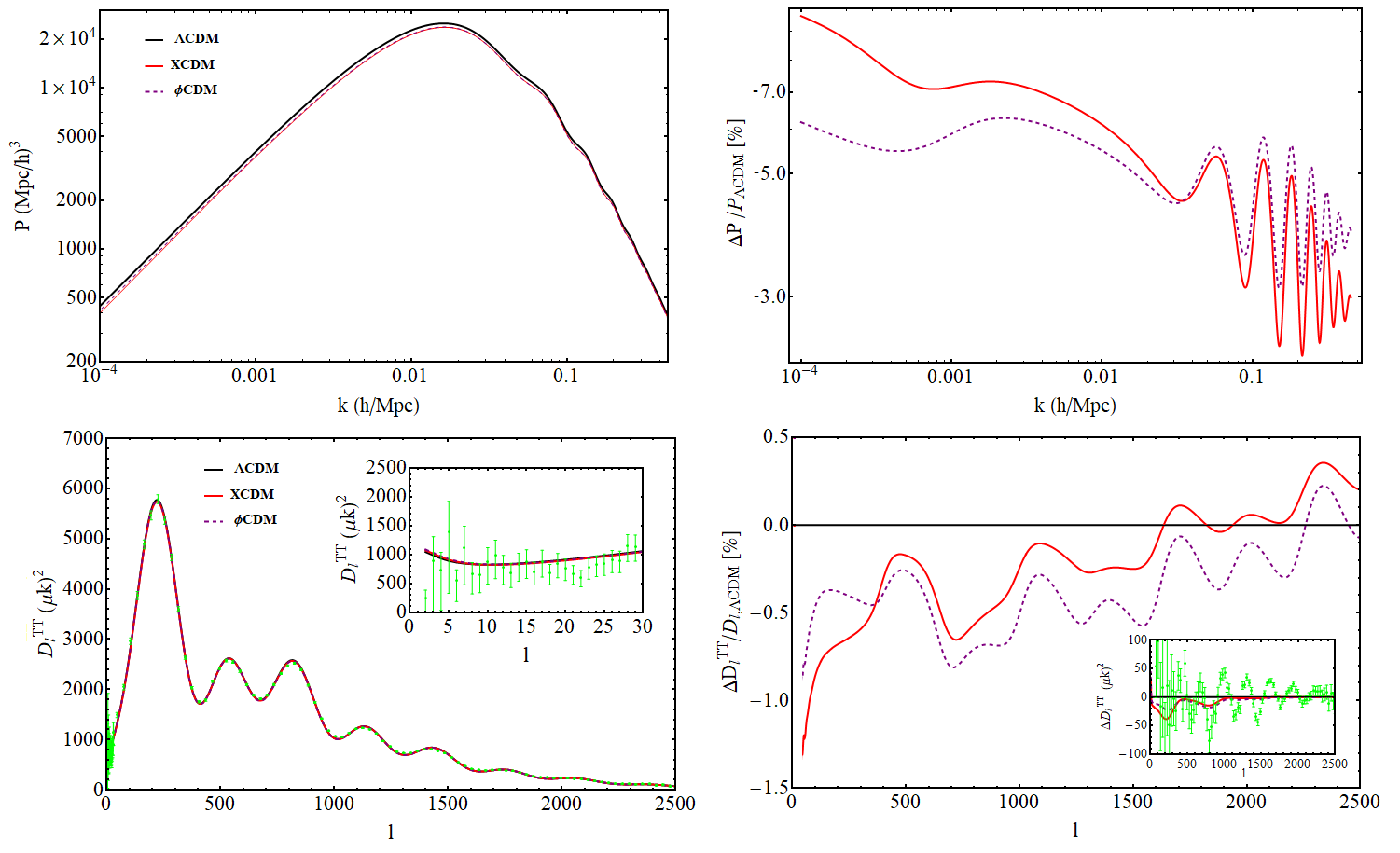}
\end{center}
\caption{\scriptsize (Upper plots) The power spectrum for the DDE models under consideration side by side with the relative (percentage) differences with respect to that of the $\CC$CDM.  We have used the best-fit values of the parameters of each model;  (Bottom plots) As before, but considering the CMB temperature anisotropies for Planck 2015 data. The inner plots show additional details including the data points with the corresponding errors in part or the whole available angular range.}
\end{figure}
%\vspace{0.3cm}
%\addtolength{\dbltextfloatsep}{+0.6cm}

%%%%%%%%%%%%%%%%%%%%%%%%%%%%%%%%%%%%%%%%%%%%%%%%%%%%%%%%%%%%%%%%%%%%%%%%%%%%%%%%%%%%%%%%%%%%%%%%%%%

%
%%%%%%%%%%%%%%%%%%%%%%%%%%%%%%%%%%%%%%%%%%%%%%%%%%%%%%%%%%%%%%%%%

%%%%%%%%%%%%%%%%%%%%%%%%%%%%%%%%%%%%%%%%%%%%%%%%%%%%%%%%%%%%%%%%%%%%%%%%%%%%%%%%%%

%%%%%%%%%%%%%%%%%%%%%%%%%%%%%%%%%%%%%%%%%%%%%%%%%%%%%%%%%%%%%%%%%%%%%%%%%%%%%%%%%%

%%%%%%%%%%%%%%%%%%%%%%%%%%%%%%%%%%%%%%%%%%%%%%%%%%%%%%%%%%%%%%%%%%%%%

\section{Bayesian evidence of DDE}
The main results of this work are synthesized in Tables 1 and 2, and in Figs. 1-10.  Here we wish to quantify the significance of these results from the statistical point of view.  This is done in detail in Fig. 10.  In what follows we explain the meaning of this figure and review some basic facts on Bayesian analysis which can be helpful at this point.  As remarked before, in the absence of BSP data the DDE signs are weak and we find consistency with previous studies. However, when we include BSP data and focus on the LSS+BAO+CMB observables (i.e. scenario DS2/BSP) the situation changes significantly.  Both models XCDM and $\phi$CDM  then consistently point to a $2.5-3\sigma$ effect.  Specifically,  the evolving EoS of the $\phi$CDM takes a value at present which lies $\sim 3\sigma$ away from the $\CC$CDM  ($w=-1$) into the quintessence region $w\gtrsim-1$ (cf. Fig. 8). The significance of these results can be further appraised by computing the Bayesian evidence, which is based on evaluating the posterior probability of a model $M$  given the data $x$ and the priors, see e.g. \cite{DEBookAmendola,AmendolaStatistics}. The relevant quantity we are looking for is the so-called Bayes factor, which does not depend on the arbitrarily assigned prior probability to the given model $M$.  If $\theta$ is the set of free parameters of such model, Bayes theorem allows us to compute the probability of measuring a distribution of values of these parameters given the measured dataset $x$ and the model $M$. It is called the posterior probability, and it is given by
\begin{eqnarray}\label{eq:BayesTheorem}
p(\theta|x,M)=\frac{p(x|\theta,M) p(\theta | M)}{p(x|M)}\,.
\end{eqnarray}
In words, it says that the posterior probability of $\theta$ is equal to the probability of the data $x$ given the parameters of the model $M$ (i.e. the likelihood  $p(x|\theta,M)$) times the prior probability of  $\theta$,  $p(\theta | M)$, divided by the probability of the data $x$  (usually in the form of a probability distribution function, PDF). Obviously the latter does not depend on the values of $\theta$. Therefore, if we normalize the posterior probability to one and  integrate over all the parameters $\theta$ on both sides of Eq.\,(\ref{eq:BayesTheorem}), the corresponding integral in the numerator on the \textit{r.h.s.}  must be equal to $p(x|M)$:
\begin{eqnarray}\label{eq:evidence}
 p(x|M) = \int d\theta\, p(x|\theta,M)\,p(\theta|M)\,.
\end{eqnarray}
This likelihood integral over all the values that can take  the parameters $\theta$ is called the marginal likelihood or evidence\,\cite{DEBookAmendola,AmendolaStatistics}.

An analogous formula to (\ref{eq:BayesTheorem}) can also be applied to estimate the posterior probability that a model $M$ is true given a measured data set $x$. Thus, following the same scheme, the posterior probability of the model given the data, $p(M|x)$, must be equal to the probability of the data given the model (irrespective of the values of $\theta$) -- i.e. the marginal likelihood (\ref{eq:evidence}) --- times the prior probability of the model, divided by the PDF of the data. Writing this relation for two cosmological models $M_i$  and  $M_j$ that are being compared, we find that the ratio of posterior probabilities of these models is equal to the ratio of model priors times a factor:
\begin{equation}
\frac{p(M_i|x)}{p(M_j|x)} =
\frac{p(M_i)}{
p(M_j)}\, B_{ij}\,.
\end{equation}
Such factor, $B_{ij} = p(x| M_i)/p(x| M_j)$,  is the so-called  Bayes factor of the model $M_i$ with respect to model $M_j$. It gives the ratio of marginal likelihoods (i.e. of evidences) of the two models. Note that it coincides with the ratio of posterior probabilities of the models only if the prior probabilities of these models are the same. This is generally assumed (``Principle of Insufficient Reason'') and hence the comparison of the two  models $M_i$ and $M_j$ is usually performed directly  in terms of $B_{ij}$.
%, which does not depend on the prior probabilities $p(M_i)$ and $p(M_j)$.

%%%%%%%%%%%%%%%%%%%%%%%%%%%%%%%%%%%%%%%%%%%%%%%%%%%%%%%%%%%%%%%%

%%%%%%%%%%%%%%%%%%%%%%%%%%%%%%%%%%%%%%%%%%%%%%%%%%%%%%%%%%%%%%%%%
%

In  the literature it has been customary to  define the Bayes information criterion (BIC) through the parameter ${\rm BIC}=\chi^2_{\rm min}+n\,\ln N$, where $\chi^2_{\rm min}$ is the minimum of $\chi^2$,  $n$ is the number of independent fitting parameters and $N$ is the total number of data points. One can show that the Bayes factor can be estimated through $B_{ij}= e^{\Delta{\rm BIC}/2}$ where $\Delta {\rm BIC}={\rm BIC}_j-{\rm BIC}_i$ is the difference between the values of BIC for models $M_j$ and $M_i$\,\cite{Kass,Burnham}, i.e.
\begin{equation}\label{eq:DeltaBIC}
\Delta {\rm BIC}=\Delta\chi^2_{\rm min}-\Delta n\,\ln N\,,
\end{equation}
%
%%%%%%%%%%%%%%%%%%%%%%%%%%%%%%%%%%%%%%%%%%%%%%%%%%%%%%%%%%%%%%%%%%%
\begin{table}[!t]
\begin{center}
\begin{tabular}{cc}
\hline
$\Delta {\rm BIC}=2\ln B_{ij}$ &$\ \ \ \ \ \ $ Bayesian evidence of model ${M}_i$ versus $M_j$ $\ \ $ \\ \hline
$0<\Delta {\rm BIC} < 2$ & {\rm weak evidence (consistency between both models)} \\
$2<\Delta {\rm BIC}< 6$ & {\rm  positive evidence} \\
$6<\Delta {\rm BIC}< 10$ & {\rm strong evidence} \\
$\Delta {\rm BIC} \geq 10$ & {\rm very strong evidence} \\
$\Delta {\rm BIC} < 0$ & {\rm counter-evidence against model $M_i$} \\
\hline
\end{tabular}
\caption{\scriptsize Conventional ranges of values of $\Delta {\rm BIC} $  used to judge the observational support for a given model $M_i$ with respect to the reference one $M_j$. See \cite{Kass,Burnham} for more details. }
\label{tab:Delta BIC}
\end{center}
\end{table}
%%%%%%%%%%%%%%%%%%%%%%%%%%%%%%%%%%%%%%%%%%%%%%%%%%%%%%%%%%%%%%%%%%%%%%%%%
%
where $ \Delta\chi^2_{\rm min}=\left(\chi^2_{\rm min}\right)_j-\left(\chi^2_{\rm min}\right)_i $ is the difference between the minimum values of  $\chi^2$ for each model,  and $\Delta n=n_i-n_j$ is the difference in the number of independent fitting parameters of $M_i$ and $M_j$, both describing the same $N$ data points. The added term to $\Delta\chi^2_{\rm min}$, i.e. $\Delta n\,\ln N$, represents the penalty assigned to the model having the largest number of parameters ($\Delta n>0$). Indeed, if $M_j$ has less parameters than $M_i$  (i.e. $n_i>n_j$) we expect that $\left(\chi^2_{\rm min}\right)_i< \left(\chi^2_{\rm min}\right)_j$ and hence $\Delta\chi^2_{\rm min}>0$. However, since $\Delta n>0$ the last term of (\ref{eq:DeltaBIC}) indeed penalizes the model with more parameters and compensates in part for its  (presumably smaller) value of $\chi^2_{\rm min}$. Defined in this way,  a positive value of the expression  (\ref{eq:DeltaBIC}) denotes that the model $M_i$ is better than model $M_j$, and the larger is $\Delta {\rm BIC}$  the better is  $M_i$ as compared to $M_j$.  Clearly, the Bayesian criterion allows a  quantitatively implementation of Occam's razor.

Even though the simple and very economic formula (\ref{eq:DeltaBIC}) is useful and has been applied in many places of the literature, see e.g. \,\cite{ApJ2017,IJMP,EPL,MNRAS2018} and references therein, it is only an approximate formula.  The exact value of $\Delta {\rm BIC}$ associated to the full Bayes factor requires to evaluate
\begin{equation}\label{eq:DeltaBICExact}
\Delta {\rm BIC}=2\,\ln B_{ij}=2\,\ln\frac{\int d\theta_i\, p(x|\theta_i,M_i)\,p(\theta_i|M_i)}{\int d\theta_j\, p(x|\theta_j,M_j)\,p(\theta_j|M_j)}\,,
\end{equation}
where $\theta_i$ and $\theta_j$ are the two sets of free parameters integrated over for each model. We refer to this (exact) value of  $\Delta {\rm BIC}$ as the full ``Bayesian evidence'' of model $M_i$ versus model $M_j$. It represents the optimal implementation of Occam's razor. Formula (\ref{eq:DeltaBIC}) provides an estimate (sometimes good, sometimes rough)  of the cumbersome expression (\ref{eq:DeltaBICExact}). The evaluation of the latter is carried out in practice using the MCMC's for the statistical analysis of the model and it can be a formidable numerical task. Fortunately it can be speeded up with the help of the recent numerical code MCEvidence\,\cite{Heavens2017}\footnote{We thank Y. Fantaye for helpful advice in the use of the numerical package MCEvidence.}. In Table 6 we collect the  ranges of values of $\Delta {\rm BIC} $ that are conventionally used in the literature to quantify the observational support of a given model $M_i$ with respect to some reference one $M_j$\,\cite{Kass,Burnham}. Consistently with the approximate formula discussed above, positive values of $\Delta {\rm BIC}$ in the indicated intervals favor the model $M_i$ over  model $M_j$. In the situation under consideration $M_j$ is the $\CC$CDM and $M_i$ is any of the considered DDE models. Small positive values near one just denote that the two models under competition are comparable  (i.e. that there is no marked preference of $M_i$ over $M_j$), but larger and positive values of $\Delta {\rm BIC} $ increase the support of $M_i$ (DDE model) versus $M_j$ ($\CC$CDM). Negative values of $\Delta {\rm BIC}$, in contrast,  would lead to the opposite conclusion, i.e. that the DDE model  is penalized as compared to the $\CC$CDM  and hence that the latter is preferred.

%
%%%%%%%% FIGURE 6 %%%%%%%%%%%%%%%%%%%%%%%%%%%%%%%%%
\begin{figure*}
\begin{center}
\label{FigLSS2}
\includegraphics[width=4.1in, height=3.1in]{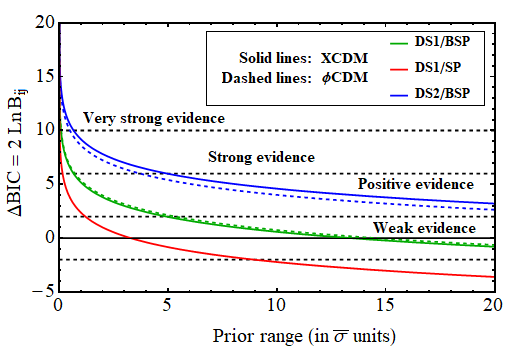}
\caption{\scriptsize  The full Bayesian evidence curves for the various models as compared to the $\CC$CDM as a function of the (flat) prior range  (in $\bar{\sigma}$ units, see text). The curves are computed using the exact evidence formula, Eq.\,(\ref{eq:DeltaBICExact}), with the data indicated in Table 1 and the first block of Table 2 (i.e. with full Planck 2015 CMB likelihood). The marked evidence ranges conform with the conventional definitions of Bayesian evidence shown in Table 6.}
\end{center}
\end{figure*}
%\addtolength{\dbltextfloatsep}{-0.5cm}

%%%%%%%%%%%%%%%%%%%%%%%%%%%%%%%%%%%%%%%%%%%%%%%%%
%

In Fig. 10 we compute the full Bayesian evidence curves as a function of  the prior range.  We assume flat priors for the parameters in each model (we have no reason to assume otherwise). Thus $p(\theta|M)$ is constant and cancels out for all  common $\theta$ in the two models, rendering $B_{ij}$  a pure ratio of marginal likelihoods. We define the unit of measure of the prior range, $\bar{\sigma}$, as half the sum of the upper and lower fitting errors of the DDE parameters in Tables 1 and 2. In Fig. 10 we rate the level of support of the DDE models as compared to the $\CC$CDM for each value of the prior range, following the standard definitions collected in Table 6\,\cite{Kass,Burnham}.

As indicated, we utilized the code MCEvidence\,\cite{Heavens2017} to compute the exact $\Delta  {\rm BIC}$ values in Fig. 10. We have actually compared these results with those obtained for the evidences computed with Gaussian (Fisher) likelihoods for the parameters and we have found that they are qualitatively similar but with non-negligible numerical differences. These were expected from the mild departures of Gaussianity from the exact distributions. The full evidence curves in Fig. 10  reconfirm the mentioned result that with the bispectrumless dataset DS1/SP  the evidence is weak. But at the same time we can recognize once more the impact of the bispectrum in the scenarios enriched with this component. Thus, it is remarkable to find a long tail of sustained positive Bayesian evidence both for the XCDM and $\phi$CDM within the scenario DS2/BSP, which corroborates the higher sensitivity of the BAO+LSS+CMB data to DDE.  The fact that with the full Planck 2015 data and with traditional models and parametrizations of the DE we can reach such significant level of evidence is encouraging. Even more reassuring is the fact that with the preliminary Planck 2018 data we can reach similar levels of evidence (cf. Table 2, second block, and also Figs. 6 and 7), which will have to be refined when the full Planck 2018 likelihood will be made public.

Last, but not least, let us emphasize one final important point of our analysis. {Our results with  BSP data show that the DDE values of $\sigma_8(0)$  (specially within the DS2/BSP scenario) are smaller than  the (bispectrumless)   $\CC$CDM value  quoted in Table 1  by roughly $2\sigma$,  thus essentially mitigating  the $\sigma_8$-tension\,\cite{Macaulay2013}. As remarked in Sect. 5 (related to our discussion  on Fig.\,3),  the $\CC$CDM value of  $\sigma_8(0)$  does also diminish with  BSP,  but it is insufficient to alleviate the $\sigma_8$-tension. The latter is  defined of course by taking as a reference the bispectrumless  $\CC$CDM result since the higher order corrections to the spectrum were never considered in previous analyses of the $\sigma_8$-tension. This is the reason  why we have compared with that value  in order to judge the performance of the DDE models in connection to such tension. } We have checked also the influence of massive neutrinos.  We find that a similar effect is possible within the $\CC$CDM (under  DS1/BSP)  for a sum of neutrinos masses of $\sum  m_\nu=  0.195\pm  0.076$ eV, leading to $\sigma_8(0)= 0.785\pm 0.014$.  Recall, however, that in all our fitting tables we have included  a light massive neutrino of  $0.06$ eV, which is the standard  assumption adopted in most analyses\,\cite{Planck} so as to preserve  the so-called normal hierarchy of neutrino masses (as required from neutrino oscillation experiments).  A solution solely based on invoking a  substantially massive neutrino scenario is thus ruled out since it clashes with the usual constraints on neutrino oscillations; specifically,  it requires unnatural fine-tuning among the neutrino masses. The DDE option is, therefore, more natural.  There may be, of course, other factors that could have an influence on this analysis, such as e.g. the possible concurrence of spatial curvature effects. When curvature is allowed to be a free parameter the constraints on DE
dynamics may weaken since we then have an additional parameter.  We have already seen a weakening of DDE evidence in the case of the CPL as compared to the XCDM, owing to the presence of one more parameter in the former case as compared to the latter.  The effects of curvature in this kind of analyses  have been studied for the main models under consideration  in\,\cite{ParkRatra_data,ParkRatraXCDM,ParkRatraPhiCDM,Ooba2018,ParkRatra2018}, and previously in\,\cite{Ooba2017,Farooq2015}. For instance, in \cite{ParkRatra_data} (see also references therein)  the authors show that in the non-flat $\Lambda$CDM model the spatial curvature also allows a decrease of the central value of $\sigma_8(0)$ when only the TT+lowP+lensing Planck 2015 CMB data is employed in the fitting analysis. When a more complete data set including SNIa+$H(z)$+BAO+LSS is considered, though, such decrease is less significant. The same pattern is found e.g. for the $\phi$CDM in \cite{ParkRatraPhiCDM}. It is nonetheless important to point out that the effect of the bispectrum signal in the BAO+LSS data sector has not been studied in the context of a spatially curved universe yet. It would be interesting to carefully assess it. Here, however, we have chosen to present our main results by placing  emphasis on the impact of the bispectrum on the study of the DDE within the simplest possible scenario, which is to assume spatially flat FLRW metric. In order to better isolate the effect, we have avoided additional considerations, except for revisiting the impossibility that massive neutrinos alone can solve these problems  in a natural way.  Owing to the considerable controversy in the literature on these matters we believe that by focusing the message we gain in clarity.  The study of additional effects can be, of course,  very interesting for future investigations, but the main focus of the current study is the impact of the bispectrum on the DDE.

%%%%%%%%%%%%%%%%%%%%%%%%%%%%%%%%%%%%%%%%%%%%%%%%%%%%%%%%%%%%%%%%%
%%%%%%%%%%%%%%%%%%%%%%%%%%%%%%%%%%%%%%%%%%%%%%%%%%%%%%%%%%%%%%%%%
%%%%%%%%%%%%%%%%%%%%%%%%%%%%%%%%%%%%%%%%%%%%%%%%%%%%%%%%%%%%%%%%%

\section{Conclusions}

In this work, we have tested the performance of cosmological physics beyond the standard or concordance $\CC$CDM model, which is built upon a rigid cosmological constant. We have shown  that the global cosmological observations can be better described in terms of models equipped with a dynamical DE component that evolves with the cosmic expansion.  Our task has focused on three dynamical dark energy (DDE) models: the  general XCDM and CPL parametrizations as well as a traditional scalar field model ($\phi$CDM), namely the original Peebles \& Ratra model based on the potential (\ref{eq:PRpotential}). We have fitted them (in conjunction with the $\CC$CDM) to the same set of cosmological data based on the observables SNIa+BAO+$H(z)$+LSS+CMB. Apart from the global fit involving all the data, we have also tested the effect of separating the expansion history data (SNIa+H(z)) from the features of CMB and the large scale structure formation data (BAO+LSS, frequently interwoven), where LSS includes both the RSD and weak lensing measurements.  We have found that the expansion history data are not particularly sensitive to the dynamical effects of the DE, whereas the BAO+LSS+CMB are more sensitive.  Furthermore, we have evaluated for the first time the impact of the bispectrum component in the matter correlation function on the dynamics of the DE .  We have done this by including BAO+LSS data that involve both the conventional power spectrum and the bispectrum. The outcome is that when the bispectrum component is not included our results are in perfect agreement with previous studies of other authors, meaning that in this case we do not find clear signs of dynamical DE. In contrast, when we  activate the bispectrum component in the BAO+LSS sector  (along with the corresponding covariance matrices) we can achieve a significant DDE signal at a confidence level of $2.5-3\,\sigma $ for the XCDM and the $\phi$CDM. We conclude that the bispectrum can be  a very useful tool to track possible dynamical features of the DE and their influence in the formation of structures in the linear regime.

A surplus of our analysis is that we have also found noticeably lower values of $\sigma_8(0)$ in the presence of the bispectrum, see the third from last row of Table 2. For a long time it has been known that there is an unexplained  `$\sigma_8$-tension' in the framework of the $\CC$CDM, which is revealed through the fact that the $\Lambda$CDM tends to provide higher values of $\sigma_8(0)$ than those obtained from RSD measurements.  Dynamical DE models, therefore, seem to provide a possible alleviation of such tension, specially when we consider the combined CMB+BAO+LSS measurements and with the inclusion of the bispectrum.

Finally, let us remark that although  we have used the full Planck 2015 CMB likelihood in our work, we have also advanced the preliminary results involving the recent Planck 2018  data under compressed CMB likelihood, still awaiting for the public appearance of the full Planck 2018 likelihood. The prospective results that we have obtained do consistently keep on favoring DDE versus a rigid cosmological constant.

To summarize, our study shows that it is possible to reach significant signs of DDE with the current data, provided we use the bispectrum in combination with the power spectrum.  The main practical conclusion that we can draw from our results is quite remarkable: the potential of the bispectrum for being sensitive to dynamical DE effects is perhaps more important than it was suspected until now. As it turns out, its more conventional application as a tool to estimate the bias between the observed galaxy spectrum and the underlying matter power spectrum may now be significantly enlarged, for the bispectrum (as the first higher-order correction to the power spectrum) might finally reveal itself as an excellent tracer of dynamical DE effects, and ultimately of the  ``fine structure'' of the DE.

\vspace{1cm}

{\bf Acknowledgements}

We are funded by projects  FPA2016-76005-C2-1-P (MINECO), 2017-SGR-929 (Generalitat de Catalunya) and MDM-2014-0369 (ICCUB). JdCP  also by a  FPI fellowship associated to FPA2016.  We are  thankful to Y. Fantaye for correspondence concerning the implementation of the package MCEvidence in the CLASS+MontePython platform used in our MCMC analysis.  We are much grateful to the anonymous Referee for the accurate examination of our work and for providing a number of insightful suggestions which have helped a lot to  improve the presentation of  our results.

\end{document}